\documentclass[amsmath,amssymb,amsfonts,twocolumn,english,showpacs,superscriptaddress,prb]{revtex4-1}
\pdfoutput=1

\usepackage[T1]{fontenc}
\usepackage[latin9]{inputenc}
\usepackage{times}
\usepackage{graphicx}
\usepackage{color}
\usepackage{mathrsfs}
\usepackage{float}
\usepackage{indentfirst}
\usepackage{babel}
\usepackage{wasysym}
\usepackage{tikz}
\usetikzlibrary{calc}

\usepackage{booktabs}
\usepackage{array, multirow}
\usepackage{subfigure}

\usepackage[colorlinks,bookmarks=false,citecolor=blue,linkcolor=blue,urlcolor=blue]{hyperref}
\usepackage[sort&compress]{natbib}
\setcitestyle{numbers,square}

 \newcommand{\beq}{\begin{equation}}
\newcommand{\eeq}{\end{equation}}
\newcommand{\beqa}{\begin{eqnarray}}
\newcommand{\eeqa}{\end{eqnarray}}
\newcommand{\bi}{\begin{itemize}}
\newcommand{\ei}{\end{itemize}}
\newcommand{\ket} [1] {\vert #1 \rangle}
\newcommand{\bra} [1] {\langle #1 \vert}
\newcommand{\braket}[1]{\langle #1 \rangle}

\newcommand{\comment}[1]{}

\newcommand{\eqnref}[1]{Eq.~(\ref{#1})}
\newcommand{\Ref}[1]{Ref.~\cite{#1}}

\def\PI{3.1415926535898}

\usetikzlibrary{decorations.markings}
\tikzset{arrowed/.style=
  {decorate,decoration={
  markings,
  mark=between positions 0.1 and 1.0 step 0.11 with {\arrowreversed{stealth}}}
  }}
\tikzset{->-/.style={decoration={
  markings,
  mark=at position .5 with {\arrow{>}}},postaction={decorate}}}
\tikzset{-<-/.style={decoration={
  markings,
  mark=at position .5 with {\arrow{<}}},postaction={decorate}}}

\makeatother
  
\graphicspath{{.}}

\begin{document}

\title{Ising Anyons in Frustration-Free Majorana-Dimer Models}

\author{Brayden Ware}
\affiliation{Department of Physics, University of California, Santa Barbara, CA 93106-6105, USA}

\author{Jun Ho Son}
\affiliation{Department of Physics and Institute for Quantum Information and Matter, California Institute of Technology, Pasadena, CA 91125, USA}

\author{Meng Cheng}
\affiliation{Station Q, Microsoft Research, Santa Barbara, CA 93106-6105, USA}

\author{Ryan V. Mishmash}
\affiliation{Department of Physics and Institute for Quantum Information and Matter, California Institute of Technology, Pasadena, CA 91125, USA}
\affiliation{Walter Burke Institute for Theoretical Physics, California Institute of Technology, Pasadena, CA 91125, USA}

\author{Jason Alicea}
\affiliation{Department of Physics and Institute for Quantum Information and Matter, California Institute of Technology, Pasadena, CA 91125, USA}
\affiliation{Walter Burke Institute for Theoretical Physics, California Institute of Technology, Pasadena, CA 91125, USA}

\author{Bela Bauer}
\affiliation{Station Q, Microsoft Research, Santa Barbara, CA 93106-6105, USA}

\begin{abstract}

Dimer models have long been a fruitful playground for understanding topological physics. Here we introduce a new class---termed
\emph{Majorana-dimer models}---wherein  bosonic dimers are decorated with pairs of Majorana modes. We find that the simplest examples
of such systems realize an intriguing,
intrinsically fermionic phase of matter that can be viewed as the product of a chiral Ising theory, which hosts deconfined non-Abelian quasiparticles,
and a topological $p_x - ip_y$ superconductor. While the bulk anyons are described by a single copy of the Ising
theory, the edge remains fully gapped.  Consequently, this phase can arise in exactly solvable, frustration-free models. We describe
two parent Hamiltonians:  one generalizes the well-known dimer model on the triangular lattice, while the other is most naturally
understood as a model of decorated fluctuating loops on a honeycomb lattice.
Using modular transformations, we show that the ground-state manifold of the latter model unambiguously exhibits all properties of the $\text{Ising} \times (p_x-ip_y)$ theory.
We also discuss generalizations with more than one Majorana mode per site, which realize phases related to Kitaev's 16-fold
way in a similar fashion.

\end{abstract}

\maketitle
 
\section{Introduction}

Since Anderson's seminal work exploring the relation of high-temperature superconductivity and resonating valence-bond physics~\cite{Anderson73, Anderson87}, dimer models
have served as a tool to explore the low-energy behavior of antiferromagnetic spin systems, where fluctuating pairs of spin-singlets are
expected to comprise the relevant degrees of freedom~\cite{rokhsar1988, Read1989, Fradkin1990}. These dimer models describe bosonic degrees of freedom on the links of the lattice
with the additional constraint that a fixed number of such dimers emanate from each lattice site. Due to the constrained nature
of the Hilbert space, dimer models afford a large degree of analytical control and have been immensely insightful in uncovering the physics
of systems beyond the standard Landau symmetry-breaking paradigm, in particular topological spin liquids~\cite{moessner2001,moessner2001-b}.

Historically, dimer configurations have often been viewed as proxies for different ways to pair neighboring spins on a lattice into singlets.
We go beyond this paradigm by introducing what we term \emph{Majorana-dimer models}: in addition to the dimer degrees of freedom on the links,
we introduce Majorana modes~\cite{Read00,kitaev2001} on the sites of the lattice. In the low-energy sector of our models, the Majorana
modes adjacent to a bond are strongly paired if a dimer is present on this bond. We will see that coupling the fermionic degrees of freedom to dimers in this way generates novel phases of matter that cannot appear in a purely bosonic model. These phases are realized as ground states of frustration-free, and,
in one of the settings, indeed exactly solvable Hamiltonians.

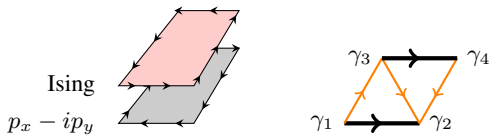
\begin{figure}
  \begin{tikzpicture}
  \draw[fill=black!20] (0,0) -- (1,0) -- (1.6,1) -- (0.8,1) -- cycle;
  \draw[arrowed] (0,0) -- (1,0) -- (1.6,1) -- (0.8,1) -- cycle;
  
  \draw[fill=red!20] (0,0.5) -- (1,0.5) -- (1.6,1.5) -- (0.8,1.5) -- cycle;
  \draw[arrowed] (0,0.5) -- (0.8,1.5) -- (1.6,1.5) -- (1,0.5) -- cycle;
  
  \node[left] at (-0.2,0) {$p_x - i p_y$};
  \node[left] at (-0.2,0.5) {Ising};
  
\begin{scope}[shift={(3,0)}]
	\draw[ultra thick, ->-] (0.0, 0.0)node[left]{$\gamma_1$}  -- (1.0, 0) node[right]{$\gamma_2$};
	\draw[thick, orange, -<-] (1.0, 0) -- ({1.5}, {sqrt(3)/2.0});
	\draw[thick, orange, ->-] (0.0, 0.0) -- ({0.5}, {sqrt(3)/2.0}) ;
	\draw[thick, orange, -<-] (1.0, 0) -- ({0.5}, {sqrt(3)/2.0});
	\draw[ultra thick, ->-] ({0.5}, {sqrt(3)/2.0}) node[left]{$\gamma_3$} -- ({1.5}, {sqrt(3)/2.0}) node[right]{$\gamma_4$};
\end{scope}

\end{tikzpicture}
  \caption{
  Left panel: Bilayer of an Ising phase and a $p_x - i p_y$ topological superconductor with opposite chirality, which together give rise to the topological
  phase discussed in this paper. This phase is characterized by three distinct topological sectors, but has a fully gapped edge.
  Right panel: The Hilbert space of Majorana-dimer models consists of bosonic dimers on the edges of the lattice and Majorana modes
  on the lattice sites. In the low-energy subspace, the Majoranas are paired according to the placement of the dimers: e.g., the fermion
  wavefunction corresponding to the dimer configuration shown is the ground state of $H_F = -i\gamma_1 \gamma_2 - i \gamma_3 \gamma_4 $. }
  \label{fig:bilayer}
\end{figure}

In the case of one Majorana mode per site of the lattice, we find realizations of Ising topological order---i.e., an Ising phase---which we substantiate both by observing the pattern of ground-state degeneracy on non-trivial manifolds
 and by computing modular matrices. Known realizations of the Ising phase, such as Kitaev's honeycomb model~\cite{Kitaev06a} or the $\nu=1$ bosonic Pfaffian fractional quantum Hall state~\cite{Greiter1991,Greiter1992}, as well as the closely related Moore-Read state for the $\nu=5/2$ plateau~\cite{moore1991}, exhibit chiral edge states (in fact required by modularity in bosonic systems~\cite{Kitaev06a}). Our models, on the contrary, support fully gapped edges. The resolution crucially relies on the fact that we are considering a \emph{fermionic} system: There is actually a ``hidden'' $p_x-ip_y$ superconductor, whose chiral Majorana edge states~\cite{Read00} exactly ``cancel'' those of the Ising phase (see Fig.~\ref{fig:bilayer}); at the same time, the $p_x-ip_y$ superconductor does not modify the universal bulk properties since it is a short-range entangled state. Therefore, our models generate an intrinsically fermionic topological phase of matter that does not exist in bosonic systems. By placing more than one Majorana mode on each site, we can construct frustration-free parent Hamiltonians for a more general class of models with gapped boundaries. For an odd number of Majorana modes per site, we realize variants of the above $\text{Ising} \times (p_x - ip_y)$ phase, while for an even number per site we realize a series of Abelian topological phases with four quasiparticles that are known from Kitaev's 16-fold way~\cite{Kitaev06a}.

Our construction starts from models of $\mathbb{Z}_2$ topological order, such as the dimer model on the triangular lattice at the Rokhsar-Kivelson point~\cite{rokhsar1988} or the toric code on the honeycomb lattice, and then couples their microscopic degrees of freedom to Majorana modes~\cite{freedman2011,freedman2011b}. 
We first explore the triangular-lattice model~\cite{moessner2001}, where Majorana modes on the lattice sites couple to the dimers in such a way that if a bond is occupied by a dimer, the complex fermion formed by the two adjacent Majoranas is, say, unoccupied. We find that there exists a local Hamiltonian---very much akin to the Rokhsar-Kivelson Hamiltonian for bosonic dimers---whose ground states are equal-weight superpositions of all dimer configurations with the corresponding Majorana configurations formed according to the above rule. The Hamiltonian is found to be frustration-free, i.e., the ground state is a simultaneous eigenstate of all terms of the Hamiltonian.
When the dimer model is in the ``resonating valence bond'' (RVB) phase, deconfined monomer excitations (i.e., sites with no emanating dimers) harbor unpaired Majorana modes, which strongly hints at the formation of an Ising-like topological phase.

In a complementary viewpoint, we describe the same phase through a model of fluctuating loops. This perspective follows a recently established paradigm of enhancing loop models by dressing the loops with one-dimensional symmetry-protected topological phases (SPT's). The approach gives a straightforward construction for symmetry-enriched versions of the corresponding loop model~\cite{Yao_unpub, LiPRB2014, chingyu2014,benzion2015}, since the ends of open strings will carry the same projective representation of the symmetry group as the edge modes of the SPT. The new ingredient here is to consider a one-dimensional fermionic topological phase---the Kitaev chain~\cite{kitaev2001}---that exhibits unpaired Majorana zero modes at the ends. Excitations formed from open strings will thus carry Majorana zero modes. By choosing the Hamiltonian such that the loops fluctuate, these excitations become deconfined and a topologically ordered Ising phase emerges. We construct a commuting-projector Hamiltonian on a Fisher lattice that exactly realizes this scenario.

A similar approach to ours, including the use of Kasteleyn orientations, was used in Ref.~\onlinecite{Tarantino2016-dn} to obtain exactly soluble parent Hamiltonians for all known fermionic
symmetry-protected topological phases with an on-site $\mathbb{Z}_2$ symmetry group.  The phases described in this
paper can be viewed as gauged versions of the phases described in~\cite{Tarantino2016-dn}, and on the Fisher lattice a duality transformation---which we discuss in more detail
in the Conclusions---can be used to establish a correspondence between the models.

The remainder of this paper is laid out as follows.
Section~\ref{sec:RK} reviews the underlying bosonic quantum dimer models that form the basis for our construction. We then discuss general properties of the Majorana-dimer model constructions in Secs.~\ref{sec:majdimers},~\ref{sec:parity}, and~\ref{sec:hamiltonian}; the precise form of the dimer dynamics is presented in Secs.~\ref{sec:loop-model} and~\ref{sec:dimer-model}. Section~\ref{sec:toporder} presents ground-state degeneracy and entanglement calculations to determine the precise nature of the topological order in these states. In Sec.~\ref{sec:8fold}, we discuss the generalizations of our model to systems with more than one Majorana mode per site, drawing on the results established in all the previous sections. Finally, we discuss our results and provide an outlook in Sec.~\ref{sec:conclusions}.

\section{Review of Bosonic Dimer Models} \label{sec:RK}

Before introducing the parent Hamiltonians of the Majorana-dimer construction, we briefly review the Rokhsar-Kivelson~\cite{rokhsar1988} Hamiltonian for
bosonic dimer models on the triangular lattice~\cite{moessner2001} and Fisher lattice~\cite{fisher1966}. In the former case,
the Hamiltonian
\begin{align} \label{eqn:bRK}
H_\text{RK}^\bigtriangleup &= \sum_p\big(-t B_p^\bigtriangleup + V C_p^\bigtriangleup\big)
\end{align}
is the sum of dimer flip and potential energy terms, represented by (for one of the plaquette orientations): 
\begin{align}
B_p^\bigtriangleup &= 
\Big|
\begin{tikzpicture}[baseline={($ (current bounding box) - (0,3.0pt) $)},scale=0.5]
	\draw[ultra thick, -] (0.0, 0.0) -- (1.0, 0);
	\draw[thick, orange, -] (1.0, 0) -- ({1.5}, {sqrt(3)/2.0});
	\draw[thick, orange, -] (0.0, 0.0) -- ({0.5}, {sqrt(3)/2.0}) ;
	\draw[thick, orange, -] (1.0, 0) -- ({0.5}, {sqrt(3)/2.0});
	\draw[ultra thick, -] ({0.5}, {sqrt(3)/2.0}) -- ({1.5}, {sqrt(3)/2.0});
\end{tikzpicture}
	\Big\rangle \Big\langle
\begin{tikzpicture}[baseline={($ (current bounding box) - (0,3.0pt) $)},scale=0.5]
	\draw[thick, orange, -] (0.0, 0.0) -- (1.0, 0);
	\draw[ultra thick,  -] (1.0, 0) -- ({1.5}, {sqrt(3)/2.0});
	\draw[ultra thick,  -] (0.0, 0.0) -- ({0.5}, {sqrt(3)/2.0});
	\draw[thick, orange, -] (1.0, 0) -- ({0.5}, {sqrt(3)/2.0});
	\draw[thick, orange, -] ({0.5}, {sqrt(3)/2.0}) -- ({1.5}, {sqrt(3)/2.0});
\end{tikzpicture} \Big| + \mathrm{h.c.}\\
C_p^\bigtriangleup &=
\Big|
\begin{tikzpicture}[baseline={($ (current bounding box) - (0,3.0pt) $)},scale=0.5]
	\draw[ultra thick, -] (0.0, 0.0) -- (1.0, 0);
	\draw[thick, orange, -] (1.0, 0) -- ({1.5}, {sqrt(3)/2.0});
	\draw[thick, orange, -] (0.0, 0.0) -- ({0.5}, {sqrt(3)/2.0}) ;
	\draw[thick, orange, -] (1.0, 0) -- ({0.5}, {sqrt(3)/2.0});
	\draw[ultra thick, -] ({0.5}, {sqrt(3)/2.0}) -- ({1.5}, {sqrt(3)/2.0});
\end{tikzpicture}
	\Big\rangle \Big\langle
\begin{tikzpicture}[baseline={($ (current bounding box) - (0,3.0pt) $)},scale=0.5]
	\draw[ultra thick, -] (0.0, 0.0) -- (1.0, 0);
	\draw[thick, orange, -] (1.0, 0) -- ({1.5}, {sqrt(3)/2.0});
	\draw[thick, orange, -] (0.0, 0.0) -- ({0.5}, {sqrt(3)/2.0}) ;
	\draw[thick, orange, -] (1.0, 0) -- ({0.5}, {sqrt(3)/2.0});
	\draw[ultra thick, -] ({0.5}, {sqrt(3)/2.0}) -- ({1.5}, {sqrt(3)/2.0});
\end{tikzpicture} \Big| + \Big|
\begin{tikzpicture}[baseline={($ (current bounding box) - (0,3.0pt) $)},scale=0.5]
	\draw[thick, orange, -] (0.0, 0.0) -- (1.0, 0);
	\draw[ultra thick,  -] (1.0, 0) -- ({1.5}, {sqrt(3)/2.0});
	\draw[ultra thick,  -] (0.0, 0.0) -- ({0.5}, {sqrt(3)/2.0});
	\draw[thick, orange, -] (1.0, 0) -- ({0.5}, {sqrt(3)/2.0});
	\draw[thick, orange, -] ({0.5}, {sqrt(3)/2.0}) -- ({1.5}, {sqrt(3)/2.0});
\end{tikzpicture}
	\Big\rangle \Big\langle
\begin{tikzpicture}[baseline={($ (current bounding box) - (0,3.0pt) $)},scale=0.5]
	\draw[thick, orange, -] (0.0, 0.0) -- (1.0, 0);
	\draw[ultra thick,  -] (1.0, 0) -- ({1.5}, {sqrt(3)/2.0});
	\draw[ultra thick,  -] (0.0, 0.0) -- ({0.5}, {sqrt(3)/2.0});
	\draw[thick, orange, -] (1.0, 0) -- ({0.5}, {sqrt(3)/2.0});
	\draw[thick, orange, -] ({0.5}, {sqrt(3)/2.0}) -- ({1.5}, {sqrt(3)/2.0});
\end{tikzpicture} \Big|
.
\end{align}
One can similarly write down these terms for the other two plaquette orientations.

This Hamiltonian is known to form a $\mathbb{Z}_2$ topologically ordered phase for $V_c<V<t$ for some critical $V_c>0$, and a
staggered phase with broken translation symmetry for $V>t$~\cite{moessner2001-b}.
At the ``RK point'' $t = V$, the ground states are exact eigenstates of each individual term of the Hamiltonian, i.e., the Hamiltonian is frustration-free. On a torus, the ground states include equal-weight superpositions of all flippable dimer configurations in each of the four topological sectors of dimers; these ground states extend into the topological phase. Additionally, there are a number of perfectly staggered dimer configurations that remain at zero energy, since they are not connected to other states by the dynamics of the Hamiltonian. These states remain ground states in the staggered phase, but are finite-energy states in the topological phase away from the RK point.
The excited states at the RK point are separated from the ground states by a gap of $\Delta \approx 0.2t$. 

One can view the dimer model as a spin model, with $S=1/2$ spins living on the edges of the lattice,
and straightforwardly translate the above Hamiltonian into spin terms; in particular, $\sigma^z_e=1$ indicates
the presence of a dimer on edge $e$ while $\sigma^z_e=-1$ corresponds to an empty bond. To enforce the dimer constraint
in the language of spins, a vertex term of the form
\begin{equation}
	J\sum_vA_v^\bigtriangleup = J\sum_v\left(\sum_{e\in v}\sigma^z_e + 4\right)^2
\end{equation}
must be added, where the sum $v$ runs over the vertices of the lattice. When $J\rightarrow \infty$, the dimer constraint is enforced strictly. 

We will also use a Rokhsar-Kivelson dimer model on the Fisher lattice~\cite{fisher1966, Fjaerestad2008-jx} obtained by decorating
the honeycomb lattice with a triangle on each site (see right panel of Fig.~\ref{fig:orientation}). The Hamiltonian is given by 
\begin{equation}
H_\text{RK}^{\varhexagon}= -t \sum_p B_p^{\varhexagon},
\end{equation}
\begin{align}
B_p^{\varhexagon} = &\left(\left|
\begin{tikzpicture}[baseline={($ (current bounding box) - (0,3.0pt) $)},scale=0.2]
    \newcommand*{\ystep}{sqrt(3)/2.0}    \foreach \typ/\aa/\bb in {1/1/1, -1/1/4, -1/4/1,  1/4/7, 1/7/4, -1/7/7}{
        \pgfmathsetmacro{\pA}{\aa-\typ}
        \pgfmathsetmacro{\pB}{\bb-\typ}
        \pgfmathsetmacro{\qA}{\aa+\typ}
        \pgfmathsetmacro{\qB}{\bb}
        \pgfmathsetmacro{\rA}{\aa}
        \pgfmathsetmacro{\rB}{\bb+\typ}
        \foreach \ai/\bi/\aj/\bj in {\pA/\pB/\qA/\qB, \qA/\qB/\rA/\rB, \rA/\rB/\pA/\pB}
        {
            \pgfmathsetmacro{\xi}{\ystep*\bi-\ystep*\ai}
            \pgfmathsetmacro{\yi}{\ai/2+\bi/2}
            \pgfmathsetmacro{\xj}{\ystep*\bj-\ystep*\aj}
            \pgfmathsetmacro{\yj}{\aj/2+\bj/2}
            \draw[thick, orange, -] (\xi, \yi) -- (\xj, \yj);
        };
   };
   \foreach \ai/\bi/\aj/\bj in {1/2/1/3, 2/5/3/6, 5/7/6/7, 2/1/3/1, 5/2/6/3, 7/5/7/6}
        {
            \pgfmathsetmacro{\xi}{\ystep*\bi-\ystep*\ai}
            \pgfmathsetmacro{\yi}{\ai/2+\bi/2}
            \pgfmathsetmacro{\xj}{\ystep*\bj-\ystep*\aj}
            \pgfmathsetmacro{\yj}{\aj/2+\bj/2}
            \draw[ultra thick, black, -] (\xi, \yi) -- (\xj, \yj);
        };
\end{tikzpicture}
	\right\rangle \left\langle
\begin{tikzpicture}[baseline={($ (current bounding box) - (0,3.0pt) $)},scale=0.2]
\newcommand*{\ystep}{sqrt(3)/2.0}
    \foreach \typ/\aa/\bb in {1/1/1, -1/1/4, -1/4/1,  1/4/7, 1/7/4, -1/7/7}{
        \pgfmathsetmacro{\pA}{\aa-\typ}
        \pgfmathsetmacro{\pB}{\bb-\typ}
        \pgfmathsetmacro{\qA}{\aa+\typ}
        \pgfmathsetmacro{\qB}{\bb}
        \pgfmathsetmacro{\rA}{\aa}
        \pgfmathsetmacro{\rB}{\bb+\typ}
        \foreach \ai/\bi/\aj/\bj in {\pA/\pB/\qA/\qB, \qA/\qB/\rA/\rB, \rA/\rB/\pA/\pB}
        {
            \pgfmathsetmacro{\xi}{\ystep*\bi-\ystep*\ai}
            \pgfmathsetmacro{\yi}{\ai/2+\bi/2}
            \pgfmathsetmacro{\xj}{\ystep*\bj-\ystep*\aj}
            \pgfmathsetmacro{\yj}{\aj/2+\bj/2}
            \draw[thick, orange, -] (\xi, \yi) -- (\xj, \yj);
        };
   };
   \foreach \ai/\bi/\aj/\bj in {1/2/1/3, 2/5/3/6, 5/7/6/7, 7/6/7/5, 6/3/5/2, 3/1/2/1}
        {
            \pgfmathsetmacro{\xi}{\ystep*\bi-\ystep*\ai}
            \pgfmathsetmacro{\yi}{\ai/2+\bi/2}
            \pgfmathsetmacro{\xj}{\ystep*\bj-\ystep*\aj}
            \pgfmathsetmacro{\yj}{\aj/2+\bj/2}
            \draw[thick, orange, -] (\xi, \yi) -- (\xj, \yj);
        };   
   \foreach \ai/\bi/\aj/\bj in {1/3/2/5, 3/6/5/7, 6/7/7/6, 7/5/6/3, 5/2/3/1, 2/1/1/2}
        {
            \pgfmathsetmacro{\xi}{\ystep*\bi-\ystep*\ai}
            \pgfmathsetmacro{\yi}{\ai/2+\bi/2}
            \pgfmathsetmacro{\xj}{\ystep*\bj-\ystep*\aj}
            \pgfmathsetmacro{\yj}{\aj/2+\bj/2}
            \draw[ultra thick, black, -] (\xi, \yi) -- (\xj, \yj);
        };
\end{tikzpicture} \right| + \mathrm{h.c.} \right)
 + \nonumber \\ 
& \left(\left|
\begin{tikzpicture}[baseline={($ (current bounding box) - (0,3.0pt) $)},scale=0.2]
    \newcommand*{\ystep}{sqrt(3)/2.0}    \foreach \typ/\aa/\bb in {1/1/1, -1/1/4, -1/4/1,  1/4/7, 1/7/4, -1/7/7}{
        \pgfmathsetmacro{\pA}{\aa-\typ}
        \pgfmathsetmacro{\pB}{\bb-\typ}
        \pgfmathsetmacro{\qA}{\aa+\typ}
        \pgfmathsetmacro{\qB}{\bb}
        \pgfmathsetmacro{\rA}{\aa}
        \pgfmathsetmacro{\rB}{\bb+\typ}
        \foreach \ai/\bi/\aj/\bj in {\pA/\pB/\qA/\qB, \qA/\qB/\rA/\rB, \rA/\rB/\pA/\pB}
        {
            \pgfmathsetmacro{\xi}{\ystep*\bi-\ystep*\ai}
            \pgfmathsetmacro{\yi}{\ai/2+\bi/2}
            \pgfmathsetmacro{\xj}{\ystep*\bj-\ystep*\aj}
            \pgfmathsetmacro{\yj}{\aj/2+\bj/2}
            \draw[thick, orange, -] (\xi, \yi) -- (\xj, \yj);
        };
   };
      \foreach \ai/\bi/\aj/\bj in {7/5/7/6}
        {
            \pgfmathsetmacro{\xi}{\ystep*\bi-\ystep*\ai}
            \pgfmathsetmacro{\yi}{\ai/2+\bi/2}
            \pgfmathsetmacro{\xj}{\ystep*\bj-\ystep*\aj}
            \pgfmathsetmacro{\yj}{\aj/2+\bj/2}
            \draw[thick, orange, -] (\xi, \yi) -- (\xj, \yj);
        };  
   \foreach \ai/\bi/\aj/\bj in {1/2/1/3, 2/5/3/6, 5/7/6/7, 2/1/3/1, 5/2/6/3, 8/4/7/5, 7/6/8/8}
        {
            \pgfmathsetmacro{\xi}{\ystep*\bi-\ystep*\ai}
            \pgfmathsetmacro{\yi}{\ai/2+\bi/2}
            \pgfmathsetmacro{\xj}{\ystep*\bj-\ystep*\aj}
            \pgfmathsetmacro{\yj}{\aj/2+\bj/2}
            \draw[ultra thick, black, -] (\xi, \yi) -- (\xj, \yj);
        };
\end{tikzpicture}
	\right\rangle \left\langle
\begin{tikzpicture}[baseline={($ (current bounding box) - (0,3.0pt) $)},scale=0.2]
\newcommand*{\ystep}{sqrt(3)/2.0}
    \foreach \typ/\aa/\bb in {1/1/1, -1/1/4, -1/4/1,  1/4/7, 1/7/4, -1/7/7}{
        \pgfmathsetmacro{\pA}{\aa-\typ}
        \pgfmathsetmacro{\pB}{\bb-\typ}
        \pgfmathsetmacro{\qA}{\aa+\typ}
        \pgfmathsetmacro{\qB}{\bb}
        \pgfmathsetmacro{\rA}{\aa}
        \pgfmathsetmacro{\rB}{\bb+\typ}
        \foreach \ai/\bi/\aj/\bj in {\pA/\pB/\qA/\qB, \qA/\qB/\rA/\rB, \rA/\rB/\pA/\pB}
        {
            \pgfmathsetmacro{\xi}{\ystep*\bi-\ystep*\ai}
            \pgfmathsetmacro{\yi}{\ai/2+\bi/2}
            \pgfmathsetmacro{\xj}{\ystep*\bj-\ystep*\aj}
            \pgfmathsetmacro{\yj}{\aj/2+\bj/2}
            \draw[thick, orange, -] (\xi, \yi) -- (\xj, \yj);
        };
   };
   \foreach \ai/\bi/\aj/\bj in {1/2/1/3, 2/5/3/6, 5/7/6/7, 7/6/7/5, 6/3/5/2, 3/1/2/1}
        {
            \pgfmathsetmacro{\xi}{\ystep*\bi-\ystep*\ai}
            \pgfmathsetmacro{\yi}{\ai/2+\bi/2}
            \pgfmathsetmacro{\xj}{\ystep*\bj-\ystep*\aj}
            \pgfmathsetmacro{\yj}{\aj/2+\bj/2}
            \draw[thick, orange, -] (\xi, \yi) -- (\xj, \yj);
        };   
   \foreach \ai/\bi/\aj/\bj in {1/3/2/5, 3/6/5/7, 6/7/8/8, 7/6/7/5, 8/4/6/3, 5/2/3/1, 2/1/1/2}
        {
            \pgfmathsetmacro{\xi}{\ystep*\bi-\ystep*\ai}
            \pgfmathsetmacro{\yi}{\ai/2+\bi/2}
            \pgfmathsetmacro{\xj}{\ystep*\bj-\ystep*\aj}
            \pgfmathsetmacro{\yj}{\aj/2+\bj/2}
            \draw[ultra thick, black, -] (\xi, \yi) -- (\xj, \yj);
        };
\end{tikzpicture} \right| + \mathrm{h.c.} \right) \ldots 
\end{align} Here the sum runs over all hexagonal plaquettes, and the $\ldots$ represents all possible (in total $32$) local flip moves involving $6$ dimers adjacent to a plaquette. This dimer model has the important property that all plaquettes are flippable in every dimer configuration, so that the potential term acts as a constant and can therefore be omitted.

Despite the apparent complexity of the Fisher-lattice dimer Hamiltonian, it admits an exceedingly simple description as a spin model.
As with the triangular lattice, the spin model is formed using spin-$1/2$ degrees of freedom to specify dimer states; however, it now suffices to place spins only on a subset of edges---the edges between triangles---since the dimer configuration on the remaining edges is completely determined when the dimer constraint is satisfied. We can thus map the model to one of dimer variables on the edges of a honeycomb lattice, which are constrained such that either $1$ or $3$ dimers emanate from each vertex of the honeycomb lattice. Using the spin-$1/2$ representation, the spin Hamiltonian in these variables can be written as 
\begin{gather}
\label{eqn:TC}
H_{\text{RK}}^{\varhexagon}  = - t \sum_p B_p^{\varhexagon} - J \sum_v A_v^{\varhexagon},
\end{gather}
where the individual terms read
\begin{align}
	A_v^{\varhexagon} &= \prod_{e \in v} \sigma^z_e
	&B_p^{\varhexagon} &= \prod_{e \in p} \sigma^x_e.
	\label{}
\end{align}
Here the $A_v^{\varhexagon}$ term enforces the dimer constraint, and the $B_p^{\varhexagon}$ term flips the dimer configuration. We also notice that this Hamiltonian is the same as the toric code on the underlying honeycomb lattice~\cite{Kitaev2003}. In fact, if we define an edge of the honeycomb lattice not occupied by a dimer ($\sigma^z=-1$) as being occupied by a string, the dimer constraint can be viewed as the closed-loop constraint for the strings. Thus, as in the toric code, the minimal excitations---violations of a single plaquette term---are dispersionless and carry energy $2t$. On a closed manifold such as a torus, the plaquette terms can only be violated in pairs, so the gap is $4t$. 

It is worth mentioning that dimer models have been generalized to describe other topological phases, such as the double semion phase~\cite{Freedman2004,levin2005,QiPRB2015, IqbalPRB2014, BuerschaperPRB2014}. The double-semion ground-state wavefunction has a simple representation in the loop basis:
\begin{equation}
	\ket{\psi_\text{DS}}=\sum_{L} (-1)^{n(L)} \ket{L}.
	\label{}
\end{equation}
Here $n(L)$ is the number of loops while $\lbrace \ket{L} \rbrace$ denotes the set of closed-loop configurations.
A similar wavefunction can be written down in the dimer representation, where the amplitude is $(-1)^{n(D)}$
with $n(D)$ being of the number of loops in the transition graph of $D$. Rokhsar-Kivelson-type models featuring the double-semion ground
state were recently found in~\Ref{QiPRB2015}.

\section{Majorana-Dimer Models}

\label{MajoranaModelsSec}

In this section, we start from the dimer models for $\mathbb{Z}_2$ topological order described in the previous section,
and describe how to couple them to fermionic degrees of freedom in a way that yields a new topologically
ordered phase. We first review the common ingredients for dressing dimer models with Majorana modes, and then discuss
the specifics of two models. We will see that dressing the dimer model on the Fisher lattice yields an exactly solvable model
with vanishing correlation length, while starting from the triangular lattice yields a much simpler, but not fully analytically solvable model.

\subsection{Majorana-Dimer Configurations} \label{sec:majdimers}

\begin{figure}
  \begin{subfigure}
 	 \centering    
    \includegraphics[width=0.5\linewidth]{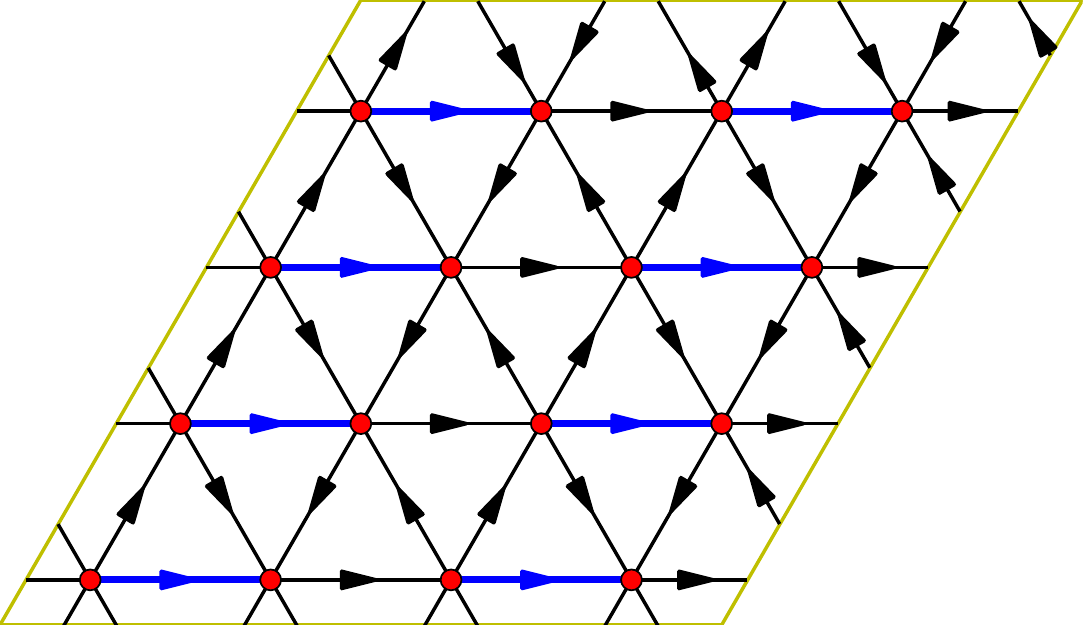} 
  \end{subfigure}
  \hspace{-1cm}
  \begin{subfigure}
  	 \centering 
    \includegraphics[angle=90, width=0.5\linewidth]{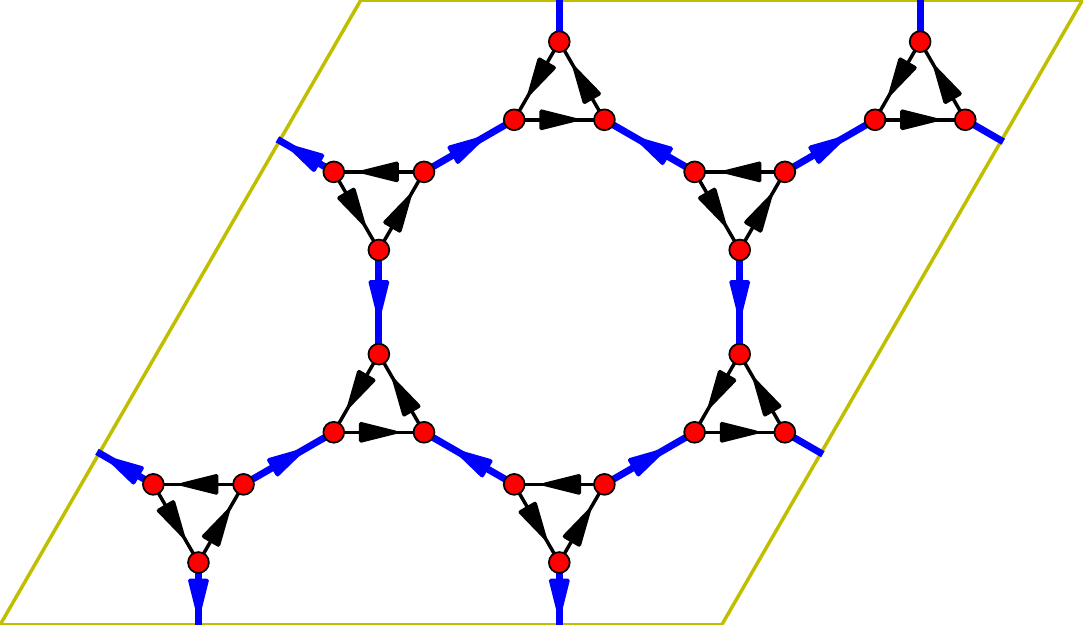} 
   \end{subfigure}
  \caption{Kasteleyn orientation (arrows) and reference dimer configuration (blue bonds) on the triangular lattice (left panel) and the Fisher lattice (right panel).}
  \label{fig:orientation}
\end{figure}

To define the Majorana-dimer models, we first associate a Majorana operator $\gamma_i$, with $\gamma_i^\dagger = \gamma_i$
and $\lbrace \gamma_i, \gamma_j \rbrace = 2 \delta_{ij}$, to each lattice site. The role of the dimers is to represent pairings of
Majorana modes into fermionic states. To uniquely define the pairings, we turn the lattice into an oriented graph by associating a direction to
each edge of the lattice. A dimer configuration is then given as a collection of oriented bonds $D = \lbrace (i,j) \rbrace$ populated by dimers.
The corresponding Majorana wavefunction $\ket{F(D)}$ is the ground state of the non-interacting Hamiltonian
\begin{equation}
H_F(D) = \sum\limits_{(i,j) \in D} i \gamma_i \gamma_j.
	\label{}
\end{equation}

In order to write down the fermionic wavefunction 
$\ket{F(D)}$, it is helpful to fix a reference set of fermion operators from which we will define a fermionic Fock space. We do that by
picking a reference dimer configuration $D_0$ on the lattice. We assign a complex fermion $f_q$ for each dimer in the reference
configuration in the following way, using the previously fixed orientation: the Majorana at the tail of the arrow is taken to be
$\gamma^A_q$ and the Majorana at the head of the arrow is taken to be $\gamma^B_q$, where 
\begin{align*}
\gamma^A_q &= f_q + f_q^\dag \\
\gamma^B_q &= i(f_q^\dag - f_q).
\end{align*}
The total dimension of this Fock space is $2^{N/2}$, where $N$ is the number of lattice sites.
Figure~\ref{fig:orientation} shows examples of reference dimer configurations, illustrated by blue bonds, for the triangular and Fisher lattices.

\begin{figure}
\includegraphics[width=0.8\linewidth]{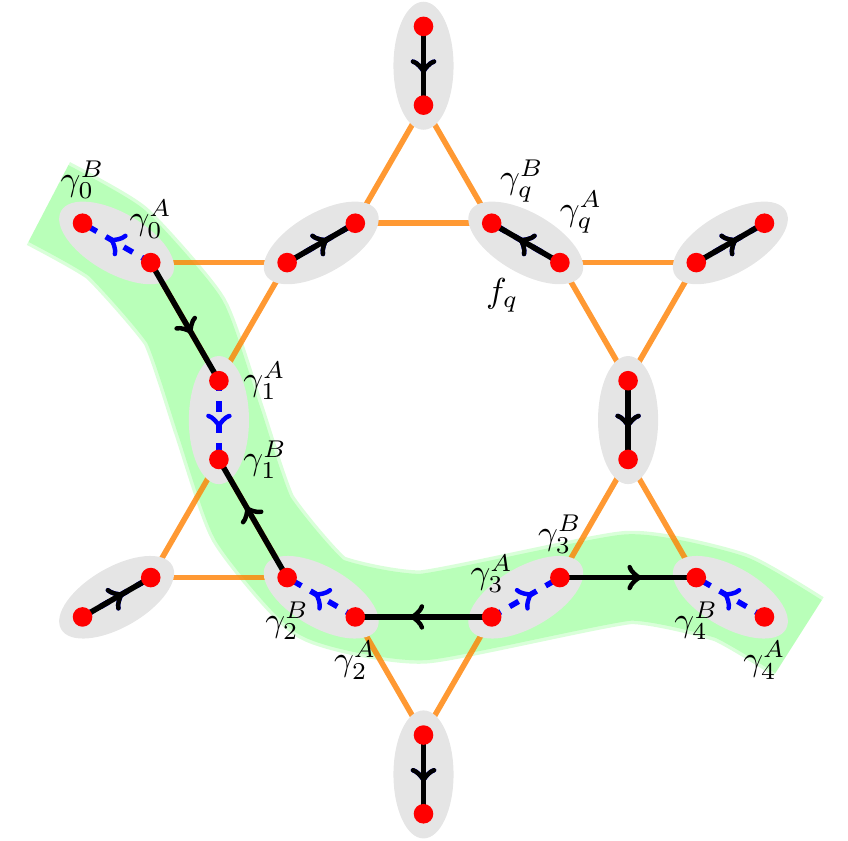}
\caption{Illustration of Majorana pairings on the Fisher lattice. The green highlighted strip illustrates part of a transition-graph loop. Away from the loop, Majorana modes pair into the reference configuration; the corresponding fermion state has each of the $f_q$ fermion unoccupied. Along the transition-graph loop, dimers are not in the reference configuration, and the Majoranas instead pair between neighboring complex fermions $f_q, f_{q'}$. The precise state of the fermions along the transition graph loop is the ground state of the Kitaev chain formed using Majoranas from neighboring sites. In the above example, this chain has the form $ h = \ldots +i\gamma^A_0 \gamma^A_1 - i \gamma^B_1 \gamma^B_2 - i \gamma^A_2 \gamma^A_3 + i \gamma^B_3 \gamma^B_4 + \ldots$. The arrow orientation on the reference edges determines the identification of the two Majoranas at each site as $\gamma^A_q$ or $\gamma^B_q$. The other arrow orientations determine the sign of the coupling between Majoranas.
}
\label{fig:kitaev}
\end{figure}

Following these rules for the definition of fermion dimers, we see that the reference dimer configuration $D_0$ corresponds to the fermion vacuum state $\ket{\mathbf{0}}$.
For some other dimer configuration $D$,
non-trivial correlations in these fermionic states arise from the fact that for dimers in $D$ that are not part of the reference state $D_0$,
the ground state of $H_F$ will pair Majoranas associated with different fermion operators $f_q$, $f_{q'}$. Relating the configuration $D$
to $D_0$ by a transition graph, we see that these non-trivial fermion pairs occur along the closed loops of the transition graph, as shown in Fig.~\ref{fig:kitaev}.
If the fermions $f_q$ of the reference configuration are viewed as the ``physical'' fermions, the coupling along such a loop resembles
the pattern of entanglement between adjacent sites in the topological phase of the Kitaev chain~\cite{kitaev2001}.

Schematically, the wavefunction we are interested in is an ``equal-weight'' superposition of Majorana-dressed dimer configurations, 
\begin{equation}
\ket{\psi} = \sum\limits_D \ket{F(D)} \ket{D}.
\label{eq:RKWF}
\end{equation}
Note that the definition of $\ket{F(D)}$ does not fix the overall phase of each $\ket{F(D)}$, so we will need to fix these phases in order to precisely define the ``equal-weight'' wavefunction.

We will also briefly consider a generalization of the wavefunction in Eq. \eqref{eq:RKWF} to include the ``double semion'' signs:
\begin{equation}
	\ket{\psi} = \sum\limits_D (-1)^{n(D)}\ket{F(D)} \ket{D}.
\label{eq:RKDSWF}
\end{equation}
Again $n(D)$ is of the number of loops in the transition graph of $D$. As we will show, this wavefunction represents a phase of matter distinct from that of Eq. \eqref{eq:RKWF}.

\subsection{Fermion Parity} \label{sec:parity}

A basic criterion for the consistency of such a fermion wavefunction is that the total fermion parity is well-defined; for
such a wavefunction to exist all superposed dimer configurations must carry the same total fermion parity. 
We will show that if the orientation for the bonds of the lattice is chosen correctly, this criterion can be met.

The fermion parity of each dimer state is described simply with a \emph{clockwise-odd rule}: if the number of arrows pointing clockwise along a transition graph loop is odd, then the fermion parity of the corresponding state is even and vice versa. (For states with multiple loops in the transition graph, the total fermion parity is determined by combining the fermion parities of each loop separately.) Appendix~\ref{app:Kasteleyn} provides a simple proof of this fact. For planar graphs, a \emph{clockwise odd} or \emph{Kasteleyn}~\cite{Kasteleyn1961} orientation can always be picked such that all transition graph loops have an odd number of clockwise-pointing arrows; choosing this orientation guarantees all Majorana-dimer states have even fermion parity. Since their introduction in Ref.~\cite{Kasteleyn1961}, these orientations have been used extensively in the study of classical, bosonic dimer models. For a lattice on a higher-genus surface such as a torus, one can only guarantee that topologically trivial transition-graph loops are clockwise odd; an orientation with this property will be considered a Kasteleyn orientation. For any such orientation, topologically non-trivial loops will be either clockwise odd or even depending only on the $\mathbb{Z}_2$ winding numbers of the loop. Thus, the wavefunction in Eq.~\eqref{eq:RKWF} will have definite fermion parity whenever the dimers in the sum belong to the same topological sector.

For the rest of this paper, we will fix a translationally invariant orientation on the torus and a reference dimer configuration for each lattice. In this case, the states of Majorana-dimers in the three topologically non-trivial sectors $(0, 1), (1, 0),$ and $(1, 1)$ have odd fermion parities, while the trivial sector $(0, 0)$ has even fermion parity. Here, the topological sectors are labelled by the parity of the winding numbers of the transition graph loops from the reference configuration.

Let us briefly discuss the effects of different orientations on the parity. To construct different classes of Kasteleyn orientations, one can flip all of the arrows on edges along one of the boundary cycles of the torus; effectively, this switches the boundary conditions for the fermions hopping across that boundary from periodic to anti-periodic. (One can also locally change the orientations by flipping the arrows surrounding a vertex, but doing so affects neither the class of orientation nor the parities.) The parity of the resulting Majorana-dimers depends on the topological sector of the dimers and on the boundary conditions as summarized in Table~\ref{tab:parity_torus}. Similar results can be derived for Kasteleyn orientations on higher genus surfaces, as detailed in Appendix~\ref{app:Kasteleyn}.

\begin{table}
	\centering
		\begin{tabular}{@{} l@{\qquad}rrrr}
         \multirow{2}{*}{\parbox[t][][b]{1cm}{Dimer \\ Sector}} & \multicolumn{4}{c}{Boundary Cond.} \\
		 & PP & PA & AP & AA\\
		\midrule
		$(0,0)$ &     $+1$ & $+1$ & $+1$ & $+1$\\
		$(1,0)$ &     $-1$ & $+1$ & $-1$ & $+1$\\
		$(0,1)$ &     $-1$ & $-1$ & $+1$ & $+1$\\
		$(1,1)$ &     $-1$ & $+1$ & $+1$ & $-1$\\
		\bottomrule
	\end{tabular}
	\caption{	The fermion parity $P_f = \pm 1$ of a Majorana-dimer state depends on the topological sector of the bosonic dimers (rows) and the boundary conditions for the fermions (columns). For the latter, `P' and `A' respectively denote periodic and antiperiodic boundary conditions. For example, PP indicates periodic boundary conditions on both cycles of the torus. }
	\label{tab:parity_torus}
\end{table}

\subsection{Phase Consistency and Ground-State Degeneracy}
\label{sec:hamiltonian}

The Majorana-dimer models that we discuss in detail below will follow the same general pattern consisting of terms that enforce the dimer constraint
around a vertex, terms that flip dimer configurations, and finally potential terms. We extend the vertex terms to not only enforce the dimer constraint, but also to force the Majorana modes to pair according to the dimer configuration and consistent with the orientation. The ground-state subspace $\mathcal{H}_r = \lbrace \ket{F(D)} \ket{D} \rbrace$ of these vertex terms is spanned by the set of allowed Majorana-dimer configurations. Additionally, we must extend the flip term $B_p$ of the dimer model to include a fermionic part $\mathcal{B}_p$ that changes the Majorana pairings along with the dimer configurations. This term provides dynamics to the Majorana-dimers and will be constructed to ensure that the ground state of the full Hamiltonian forms an equal-weight superposition of dimers as in Eq.~\eqref{eq:RKWF}.

Let $\mathbf{B}_p \equiv B_p \mathcal{B}_p$ denote the combined boson-fermion flip term. (Here and below we use bold font for those operators that act on both the bosonic and fermionic Hilbert spaces.)
Because of the bosonic part $B_p$, matrix elements $\bra{F(D')} \bra{D'}\mathbf{B}_p \ket{F(D)} \ket{D}$ within the restricted subspace $\mathcal{H}_r$ are non-zero only when dimer configurations $D$ and $D'$ differ by a single plaquette flip:
\begin{equation}
  \bra{F(D')}\bra{D'}\mathbf{B}_p \ket{F(D)} \ket{D} = e^{i \varphi_{p,D}} \delta_{D', D_p},
\label{eq:Bp}
\end{equation}
where dimer configurations $D$ and $D_p$ differ by flipping the plaquette $p$.   
Importantly, the fermionic part of the flip term contributes phases $e^{i \varphi_{p, D}}$, which are absent in the bosonic dimer model. 
We can characterize the effect of these phases by examining the matrix elements of the Hamiltonian in the reduced Hilbert space $\mathcal{H}_r$,
\begin{align}
  h_{D D'} &\equiv \langle F(D') | \langle D' |  H |F(D)\rangle |D\rangle  \nonumber \\
  &= V n(D) \delta_{D',D} - t\sum_{p} e^{i \varphi_{p,D}} \delta_{D',D_p}.
           \label{eq:hDD}
\end{align} 
All of the off-diagonal matrix elements of $h$ are generated by $\mathbf{B}_p$, while the diagonal elements of $h$ are the same as in the bosonic dimer model, i.e., the coefficient $V$ times the number of flippable plaquettes $n(D)$. (In the Fisher-lattice dimer model, $V$ appears as an overall constant and can be dropped.)

Note that $h_{DD'}$ can be viewed as a (possibly non-local) Hamiltonian acting on bosonic dimers without the accompanying Majoranas.  
The spectrum of our Hamiltonian in the restricted Hilbert space is the same as the spectrum of $h_{D D'}$ and is clearly unaffected by a redefinition $\ket{F(D)} \to e^{i \phi_D} \ket{F(D)}$. 
If such a redefinition could be made to satisfy $h_{D D'} = -t$ for all $D, D'$ that differ by a single plaquette flip, then 
the spectrum of the Majorana-dimer model in the restricted Hilbert space would be \emph{identical} to the bosonic dimer model for arbitrary $t, V$. In that case we say that the $h_{D D'}$ matrix is unfrustrated.

In the following models, we find that $h_{D D'}$ is indeed unfrustrated in systems with open boundary conditions---guaranteeing the existence of a choice of phases for $\ket{F(D)}$ where the ground state is given by Eq.~\eqref{eq:RKWF} at the RK point. As detailed in Appendix~\ref{sec:obc}, this choice is equivalent to adopting conventions for $\ket{F(D)}$ where the overlaps $\braket{\mathbf{0}|F(D)}$ are always real and positive.

The situation is more subtle, however, on closed manifolds. On a torus we find that non-trivial phases
\begin{equation}
\Theta_{\{D_k\}} = \text{ Arg} \left( h_{D_1 D_2} h_{D_2 D_3} \ldots h_{D_L D_1} \right)
\end{equation}
can be generated by a sequence of dimer flip moves that start and end with the same dimer configuration. These phases cannot be removed by any redefinition 
$\ket{F(D)} \to e^{i \phi_D} \ket{F(D)}$ and thus frustrate the hopping.
Remarkably, for each of our models, these non-trivial phases are occur only in one of the four topologically distinct sectors, namely the $(0,0)$ sector.
As a result, the minimum energy for the $(0, 0)$ sector is greater than zero, while the other three sectors admit zero-energy ground states that are equal-weight superpositions of Majorana-dimer configurations.

Thus, while the bosonic quantum dimer models on these lattices have $4$ degenerate ground states formed by superpositions of dimer configurations in each topological sector, the dynamics of the Majorana-dimers instead lead to three fermion-parity-odd ground states corresponding to superpositions of Majorana-dimer configurations in the $(0, 1), (1, 0),$ and $(1, 1)$ sectors, with a finite gap to the $(0,0)$ sector as well as to all other states. 
This reduction in ground-state degeneracy from $4\rightarrow 3$ is essential for reconciling the anyonic content of the topological order for our Majorana-dimer models discussed below. We also emphasize that the phases $\Theta_{\{D_k\}}$ can only be reproduced in the pure bosonic dimer model non-locally, while they appear from purely local dynamics in the Majorana-dimer models.  

The next two sections explain the precise form of the dynamics for a commuting-projector model on the Fisher lattice and a frustration-free model on the triangular lattice. We will then use the ground state(s) on the Fisher lattice to diagnose the topological order.

\subsection{Majorana Loop Model on a Honeycomb Lattice} \label{sec:loop-model}
Our construction on the Fisher lattice offers the advantage of having a vanishing correlation length and therefore being most amenable to both analytical and numerical methods.  
As reviewed in Sec.~\ref{sec:RK}, 
the quantum dimer model on this lattice is equivalent to a $\mathbb{Z}_2$ toric code on the associated honeycomb lattice. Dimer configurations on the Fisher lattice are in one-to-one correspondence with loops on the honeycomb lattice, as illustrated in Fig.~\ref{fig:kitaev}. We will therefore formulate the model as a decorated toric-code model, where the ground-state wavefunction is an equal-weight superposition of closed loops dressed by Kitaev chains.

The fermionic degrees of freedom for the Majorana-dimer model on this lattice consist of one complex fermion $f_e$ on each edge $e$ of the honeycomb lattice, i.e., the complex fermions lie on the sites of a Kagome lattice. We split each fermion into two Majoranas via $f_e=\frac{1}{2}(\gamma_e^A+i\gamma_e^B)$. The Majoranas now form a Fisher lattice, and we take $\gamma_e^A$ to sit at the tail of the edge's arrow. The Kasteleyn orientation in the right panel of Fig.~\ref{fig:orientation} is such that all $\gamma^{A/B}$ are naturally associated with $A/B$ sublattices. In the reference state all fermionic modes are empty $f_e^\dag f_e=0$. We pair up Majoranas according to the corresponding dimer configuration following the prescription sketched in Fig.~\ref{fig:kitaev} and described in the previous subsections.

We now define a frustration-free Hamiltonian whose ground states are given by the Majorana-loop wavefunctions introduced above. The Hamiltonian
follows the same structure as the toric code Hamiltonian [Eq.~\eqref{eqn:TC}] in that one term penalizes configurations that violate the loop or Majorana-pairing constraints,
while the second term ensures that the loops fluctuate and the ground state is an equal-weight superposition of all valid configurations.
The terms that enforce the constraints are given as the following projectors on the edges and the vertices of the honeycomb lattice:
\begin{equation}
	\begin{gathered}
		A_{1,v}^{\varhexagon}= \frac{1}{2}\left(1+\prod_{e\in v}\sigma^z_e\right)\\
	\mathbf{A}_{2,v}^{\varhexagon}=\sum_{\substack{e,e'\in v\\ e\neq e'}}\frac{1-\sigma^z_e}{2}\frac{1-\sigma^z_{e'}}{2}\frac{1+is_{e,e'}\gamma_{e}^{\lambda}\gamma_{e'}^{\lambda}}{2}\\
	\mathbf{A}_e^{\varhexagon}=  \frac{1-\sigma^z_e}{2}\frac{1+i\gamma_e^A\gamma_e^B}{2}.\\
	\end{gathered}
	\label{}
\end{equation}
Here $\lambda(v)$ in $\mathbf{A}_{2,v}^{\varhexagon}$ indicates the sublattice type of the vertex $v$; $A_{1,v}^{\varhexagon}$ enforces the loop constraint while $\mathbf{A}_{2,v}^{\varhexagon}$, $\mathbf{A}_e^{\varhexagon}$ enforce the Majorana-pairing constraints; and $s_{e,e'} = \pm 1$ encode the Kasteleyn orientation. 

We then need the plaquette term to make the loops fluctuate, which in the purely bosonic model is achieved by the first term in \eqnref{eqn:TC}.
However, in the present case, we also need to change the Majorana pairings accordingly. This will be implemented by a fermionic plaquette operator $\mathcal{B}_p^{\varhexagon}$, which only involves the Majoranas along the transition loop. We first define $\mathcal{B}_p^{\varhexagon}$ through its matrix elements between states in the Fock space of valid Majorana-dimers corresponding to dimer configurations $D$ and $D_p$ that are related by flipping plaquette $p$ [all other matrix elements will vanish when we include the contribution from the bosonic dimers and thus do not need to be specified; see Eq.~\eqref{eq:fullham} below]:
\begin{eqnarray}
	\langle F(D_p)|\mathcal{B}_p^{\varhexagon}|F(D)\rangle=
		\frac{\langle F(D_p)|F(D)\rangle}{|\langle F(D_p)|F(D)\rangle|},
	\label{eqn:overlap_Bp}
\end{eqnarray}
It is easy to see that $\mathcal{B}_p^{\varhexagon}$ is Hermitian and satisfies $\big(\mathcal{B}_p^{\varhexagon}\big)^2=1$ when acting in the restricted Hilbert space. Since $D$ and $D_p$ only differ locally, one can expect that such matrix elements can be generated by local operators.

Our specific choice of the $\mathcal{B}_p^{\varhexagon}$ operator moves Majoranas along the transition loop using a series of braids. Let us label the Majoranas along the transition loop as $\gamma_1, \gamma_2, \dots, \gamma_n$, in counterclockwise order. Here the only requirement is that $\gamma_1$ should be any of the Majoranas on the edges of the plaquette. We define $s_{i,i+1}=\pm 1$ according to the Kasteleyn orientation on the dimer connecting $\gamma_i$ and $\gamma_{i+1}$ (so that $is_{i,i+1}\gamma_i\gamma_{i+1}=1$ either before or after the move), and generally $s_{ij}=s_{i,i+1}\cdots s_{j-1, j}$ for $1\leq i<j\leq n$. 

We can now define the fermionic part of the plaquette operator as
\begin{equation}
	\mathcal{B}_p^{\varhexagon}\ket{F(D)}=U_{1,2n-1}\cdots U_{1,5}U_{1,3}\ket{F(D)}.
	\label{eqn:Bpfisher}
\end{equation}
Here, the unitary operator $U_{ij}$ exchanges two Majoranas $\gamma_i$ and $\gamma_j$:
\begin{gather}
U_{ij}= \frac{1+s_{ij}\gamma_i\gamma_j}{\sqrt{2}} \\
U_{ij}\gamma_i U_{ij}^\dag =s_{ij}\gamma_j \ \ \ \ \ \  U_{ij}\gamma_j U_{ij}^\dag =-s_{ij}\gamma_i.
\end{gather}
We show in Appendix~\ref{sec:proof-plaq} that the matrix elements of $\mathcal{B}_p^{\varhexagon}$ on the Majorana-dimer subspace indeed satisfy Eq.~\eqref{eqn:overlap_Bp}, and are therefore independent of the position of the starting Majorana $\gamma_1$ on the transition loop. As explained in Appendix~\ref{sec:proof-plaq}, the form of the $\mathcal{B}_p^{\varhexagon}$ operator is not unique;
however, any choice generates the same matrix elements given in Eq.~\eqref{eqn:overlap_Bp}.

In the restricted Hilbert space $\mathcal{H}_r$, the full plaquette operator $\mathbf{B}_p^{\varhexagon}$ acts as
\begin{equation}
	\begin{split}
	\mathbf{B}_p^{\varhexagon}\ket{\{\sigma^z\}}\ket{F}&=B_p^{\varhexagon}\ket{\{\sigma^z\}}\otimes \mathcal{B}_p^{\varhexagon} \ket{F}\\
	&=\left(\prod_{e\in p}\sigma_e^x\right)\ket{\{\sigma^z\}}\otimes  \mathcal{B}_p^{\varhexagon}\ket{F}.
	\end{split}
	\label{}
\end{equation}
The most important properties of these operators are that they commute with each other within the restricted Hilbert space,
\begin{equation}
	\mathbf{B}_p^{\varhexagon}\mathbf{B}_{p'}^{\varhexagon}=\mathbf{B}_{p'}^{\varhexagon}\mathbf{B}_p^{\varhexagon},
	\label{}
\end{equation}
and moreover that each squares to the identity, $(\mathbf{B}_p^{\varhexagon})^2=1$, as noted earlier. The proof of the commutation relation is rather technical, so we refer interested readers to Appendix~\ref{sec:proof-plaq} for details.

As described thus far, the Hamiltonian is frustration-free, i.e., the ground state is a simultaneous eigenstate of all terms. Furthermore, since all terms commute on
the restricted subspace $\mathcal{H}_r$, the stronger condition of a commuting-projector Hamiltonian in the full Hibert space can be obtained by conjugating the plaquette flip term
with appropriate projectors into $\mathcal{H}_r$.
In summary, the full Hamiltonian for this model is
\begin{equation}
\begin{gathered}
 H = -J_v\sum_v \left( A_{1,v}^{\varhexagon} + \mathbf{A}_{2,v}^{\varhexagon} \right) -J_e \sum_e \mathbf{A}_e^{\varhexagon} \\
- t \sum_p \mathbf{B}_p^{\varhexagon} \prod_{v \in p} A_{1,v}^{\varhexagon} \mathbf{A}_{2,v}^{\varhexagon} \prod_{e \in p} \mathbf{A}_e^{\varhexagon}.
\end{gathered}
\label{eq:fullham}
\end{equation}

We can also write down a Hamiltonian for the double-semion version of the wavefunction given in Eq.~\eqref{eq:RKDSWF}, by modifying the bosonic part $B_p^{\varhexagon}$ of the plaquette term to the following~\cite{levin2005}:
\begin{equation}
	B_p^{\varhexagon}=\prod_{e\in p}\sigma^x_e \cdot i^{\sum_{l\in{ p\text{ legs}}} \frac{1-\sigma_l^z}{2}}.
	\label{}
\end{equation}
Since this affects only the bosonic part, all properties related to the coupling to Majoranas are preserved.

\subsubsection{Spectrum}
Since all $\mathbf{B}_p^{\varhexagon}$ commute with each other, they can be simultaneously diagonalized. The eigenstates can then be labeled by the list of eigenvalues $b_p=\pm 1$ of $\mathbf{B}_p^{\varhexagon}$ for all $p$, with the energy $E=-\sum_p b_p$. The ground state(s) would naively correspond to $b_p=1$, and all we need to do is to determine the ground-state degeneracy. However, there are additional constraints among the plaquette operators that must be fully taken into account to correctly count the ground states---which turn out to depend on the topology of underlying manifold and the global fermion parity.

First of all, let us consider placing the model in a disk. In this case, there are no additional relations between the plaquette operators, and there are no topologically nontrivial loop configurations. The restricted Hamiltonian $h_{D D'}$ considered in Sec.~\ref{sec:hamiltonian} is unfrustrated. So the ground state is unique, with a completely gapped spectrum.

Now consider the system on a torus. Loop states are divided into four topological sectors,
distinguished by the parity of the winding number around the two non-trivial cycles.
As we have discussed in Sec.~\ref{sec:parity}, all Majorana states in a fixed topological sector of loops with given boundary conditions have the same fermion parity $P_f$.
In particular, for periodic boundary conditions, there are three degenerate ground states $(1,0),(0,1),(1,1)$ all having odd global fermion parity, and the $(0,0)$
sector has an even fermion parity. Interestingly, we observe that the only frustrating phases in the restricted Hamiltonian $h_{D D'}$ arise from sequences of dimer flips in the $(0, 0)$ sector that flip every plaquette once. This can be translated into the following global constraint:
\begin{equation}
	\prod_p \mathbf{B}_p^{\varhexagon} =-P_f.
	\label{eqn:productBp}
\end{equation}
In the three fermion-parity odd sectors with $P_f=-1$ it is possible to have ${b}_p^{\varhexagon}=1$ for all plaquettes $p$ simultaneously. This reproduces the expected three-fold ground-state degeneracy. In the even-parity sector $(0,0)$, at least one of the ${b}_p^{\varhexagon}$ must be $-1$; superpositions of dimers in the $(0,0)$ sector form the lowest excited states of the model, with a degeneracy of $N_p$, since there are $N_p$ different ways to violate exactly one plaquette. We can also interpret this result in terms of the quasiparticle excitations of the model: in the restricted Hilbert space on a closed manifold, a single fermion excitation is always bound to a plaquette flip ${b}_p^{\varhexagon}=-1$.

\subsection{Majorana-Dimer Model on a Triangular Lattice} \label{sec:dimer-model}
The Majorana loop model introduced in the previous section, albeit exactly solvable, has quite complicated plaquette terms. 
In this section we describe a triangular-lattice Majorana-dimer model that exhibits much simpler plaquette terms and naturally generalizes the bosonic dimer Hamiltonian in Eq.~\eqref{eqn:bRK}. First we need to construct local terms in the Hamiltonian that favor the correct Majorana pairing for a given dimer configuration. These are similar to the vertex and edge terms in the honeycomb lattice model and will not be repeated here.
In the following we mainly consider the limit where these binding terms are dominant, allowing us to work within the restricted Hilbert space $\mathcal{H}_r$.

The potential term is diagonal in the dimer basis and does not involve Majorana operators; this piece therefore takes exactly the same form as in Eq.~\eqref{eqn:bRK}. The flip term must, however, once again modify the bosonic dimers along with the accompanying Majoranas, which can be accomplished by supplementing the bosonic flip operator with braid matrices as follows: \begin{equation}
		\mathbf{B}_p^\bigtriangleup=e^{i\theta_p}
		\begin{cases}
		&\Big|
\begin{tikzpicture}[baseline={($ (current bounding box) - (0,3.0pt) $)},scale=0.5]
	\draw[ultra thick, -] (0.0, 0.0) -- (1.0, 0) node[right]{\scalebox{0.7}{1}};
	\draw[thick, orange, -] (1.0, 0) -- ({1.5}, {sqrt(3)/2.0});
	\draw[thick, orange, -] (0.0, 0.0) -- ({0.5}, {sqrt(3)/2.0}) ;
	\draw[thick, orange, -] (1.0, 0) -- ({0.5}, {sqrt(3)/2.0});
	\draw[ultra thick, -] ({0.5}, {sqrt(3)/2.0}) node[left]{\scalebox{0.7}{2}} -- ({1.5}, {sqrt(3)/2.0});
\end{tikzpicture}
	\Big\rangle
\Big\langle
\begin{tikzpicture}[baseline={($ (current bounding box) - (0,3.0pt) $)},scale=0.5]
	\draw[thick, orange, -] (0.0, 0.0) -- (1.0, 0);
	\draw[ultra thick,  -] (1.0, 0) node[right]{\scalebox{0.7}{1}}-- ({1.5}, {sqrt(3)/2.0});
	\draw[ultra thick,  -] (0.0, 0.0) -- ({0.5}, {sqrt(3)/2.0}) node[left]{\scalebox{0.7}{2}};
	\draw[thick, orange, -] (1.0, 0) -- ({0.5}, {sqrt(3)/2.0});
	\draw[thick, orange, -] ({0.5}, {sqrt(3)/2.0}) -- ({1.5}, {sqrt(3)/2.0});
\end{tikzpicture}
\Big|\otimes U_{12}\\
&\Big|
\begin{tikzpicture}[baseline={($ (current bounding box) - (0,3.0pt) $)},scale=0.5]
	\draw[ultra thick, -] (0.0, 0.0) -- (-1.0, 0) node[left]{\scalebox{0.7}{2}};
	\draw[thick, orange, -] (-1.0, 0) -- ({-1.5}, {sqrt(3)/2.0});
	\draw[thick, orange, -] (0.0, 0.0) -- ({-0.5}, {sqrt(3)/2.0}) ;
	\draw[thick, orange, -] (-1.0, 0) -- ({-0.5}, {sqrt(3)/2.0});
	\draw[ultra thick, -] ({-0.5}, {sqrt(3)/2.0}) node[right]{\scalebox{0.7}{1}} -- ({-1.5}, {sqrt(3)/2.0});
\end{tikzpicture}
	\Big\rangle
\Big\langle
\begin{tikzpicture}[baseline={($ (current bounding box) - (0,3.0pt) $)},scale=0.5]
	\draw[thick, orange, -] (0.0, 0.0) -- (-1.0, 0);
	\draw[ultra thick,  -] (-1.0, 0) node[left]{\scalebox{0.7}{2}}-- ({-1.5}, {sqrt(3)/2.0});
	\draw[ultra thick,  -] (0.0, 0.0) -- ({-0.5}, {sqrt(3)/2.0}) node[right]{\scalebox{0.7}{1}};
	\draw[thick, orange, -] (-1.0, 0) -- ({-0.5}, {sqrt(3)/2.0});
	\draw[thick, orange, -] ({-0.5}, {sqrt(3)/2.0}) -- ({-1.5}, {sqrt(3)/2.0});
\end{tikzpicture}
\Big|\otimes U_{12}\\
&\Big|
\begin{tikzpicture}[baseline={($ (current bounding box) - (0,3.0pt) $)},scale=0.5]
	\draw[ultra thick, -] (0.0, 0.0) -- (0.5, {sqrt(3)/2}) node[right]{\scalebox{0.7}{1}};
	\draw[thick, orange, -] (0.5, {sqrt(3)/2})  -- (0,{sqrt(3)});
	\draw[thick, orange, -] (0.5, {sqrt(3)/2}) -- (-0.5,{sqrt(3)/2});
	\draw[thick, orange, -] (0.0, 0.0) -- ({-0.5}, {sqrt(3)/2.0}) ;
	\draw[ultra thick, -] ({-0.5}, {sqrt(3)/2.0}) node[left]{\scalebox{0.7}{2}}-- (0, {sqrt(3)});
\end{tikzpicture}
\Big\rangle
\Big\langle
\begin{tikzpicture}[baseline={($ (current bounding box) - (0,3.0pt) $)},scale=0.5]
	\draw[thick, orange, -] (0.0, 0.0) -- (0.5, {sqrt(3)/2}) ;
	\draw[ultra thick, -] (0.5, {sqrt(3)/2}) node[right]{\scalebox{0.7}{1}} -- (0,{sqrt(3)});
	\draw[thick, orange, -] (0.5, {sqrt(3)/2}) -- (-0.5,{sqrt(3)/2});
	\draw[ultra thick,  -] (0.0, 0.0) -- ({-0.5}, {sqrt(3)/2.0}) node[left]{\scalebox{0.7}{2}};
	\draw[thick, orange, -] ({-0.5}, {sqrt(3)/2.0}) -- (0, {sqrt(3)});
\end{tikzpicture}
\Big|\otimes U_{12} \\
&\Big|
\begin{tikzpicture}[baseline={($ (current bounding box) - (0,3.0pt) $)},scale=0.5]
	\draw[ultra thick, -] (0.0, 0.0) -- (0.5, {sqrt(3)/2}) node[right]{\scalebox{0.7}{1}};
	\draw[thick, orange, -] (0.5, {sqrt(3)/2})  -- (0,{sqrt(3)});
	\draw[ultra thick, blue, -] (0.5, {sqrt(3)/2}) -- (-0.5,{sqrt(3)/2});
	\draw[thick, orange, -] (0.0, 0.0) -- ({-0.5}, {sqrt(3)/2.0}) ;
	\draw[ultra thick, -] ({-0.5}, {sqrt(3)/2.0}) node[left]{\scalebox{0.7}{2}}-- (0, {sqrt(3)});
\end{tikzpicture}
\Big\rangle
\Big\langle
\begin{tikzpicture}[baseline={($ (current bounding box) - (0,3.0pt) $)},scale=0.5]
	\draw[thick, orange, -] (0.0, 0.0) -- (0.5, {sqrt(3)/2}) ;
	\draw[ultra thick, -] (0.5, {sqrt(3)/2}) node[right]{\scalebox{0.7}{1}} -- (0,{sqrt(3)});
	\draw[ultra thick, blue, -] (0.5, {sqrt(3)/2}) -- (-0.5,{sqrt(3)/2});
	\draw[ultra thick,  -] (0.0, 0.0) -- ({-0.5}, {sqrt(3)/2.0}) node[left]{\scalebox{0.7}{2}};
	\draw[thick, orange, -] ({-0.5}, {sqrt(3)/2.0}) -- (0, {sqrt(3)});
\end{tikzpicture}
\Big|\otimes U_{12}.
\end{cases}+\text{h.c.}
\label{eqn:dimer-flip}
\end{equation} Here $U_{12}=(1+s_{12}\gamma_1\gamma_2)/\sqrt{2}$, with the $\gamma_{1,2}$ operators labeled as above and $s_{12}$ defined by the Kasteleyn orientation. The phase factors $e^{i\theta_p}$ are explained below --- for now we simply note that $e^{i \theta_p} = e^{i\pi/4}$ for plaquettes whose interior bond coincides with the position of a reference dimer [blue bonds in Fig.~\ref{fig:orientation}],
while for all other plaquettes $e^{i \theta_p} = 1$. One can check that the braid operators indeed give the desired Majorana pairings.

The full Hamiltonian, constructed analogously to the Fisher-lattice model of Eq.~\eqref{eq:fullham}, reads
\begin{equation}
\begin{gathered}
 H = J_v\sum_v A_{v}^\bigtriangleup - J_e \sum_e \mathbf{A}_e^\bigtriangleup 
- \sum_p \big(t \mathbf{B}_p^\bigtriangleup - V C_p^\bigtriangleup\big),
\end{gathered}
\label{eq:fullhamtri}
\end{equation}
where 
\begin{equation}
	\mathbf{A}_e^\bigtriangleup =
	\frac{1-\sigma^z_e}{2}\frac{1+ is_{ij} \gamma_i\gamma_j}{2}
\end{equation}
is the Majorana-dimer projector at an edge $e$ that connects vertices $i, j$.
There are two minor differences from Eq.~\eqref{eq:fullham}: the additional potential term $C_p$ from the bosonic dimer model on the triangular lattice is needed, and additional projectors previously tacked onto the flip term $\mathbf{B}_p$ no longer appear. This latter choice is made to simplify the Hamiltonian and does not affect the existence of frustration-free ground states; however, neighboring $\mathbf{B}_p$ terms do not commute in this model independent of whether the additional projectors are present or absent. 

We include the phase factors $e^{i\theta_p}$ in Eq.~\eqref{eqn:dimer-flip} for the following reason: As explained in Appendix~\ref{sec:obc}, with open boundary conditions (e.g., on a disk) the Majorana-dimer flip term as written is \emph{unfrustrated} in the sense described in Sec.~\ref{sec:hamiltonian}. 
It follows that in open boundary conditions the spectrum of the Majorana-dimer model within the restricted subspace is identical to that of the bosonic dimer model for any $t/V$. 
This mapping allows us to directly import known results for the bosonic dimer model to the present case. For example, at the RK point $t=V$, the ground state wavefunction is the equal-weight superposition of all Majorana-dimer states given in Eq.~\eqref{eq:RKWF}. Additionally, we see that no gapless edge modes are present, as found for the commuting-projector Majorana loop model discussed previously.

While these rigorous analytical statements do not simply extend to the torus, numerical evidence from small clusters strongly suggests that the topological ground-state degeneracy also matches that of the Majorana loop model. We have checked numerically on $4 \times 4$, $6 \times 4$ and $4 \times 6$ lattices with periodic boundary conditions that the Majorana-dimer model remains unfrustrated in the $(1, 0), (0, 1), (1, 1)$ sectors, but not in the $(0, 0)$ sector. [More precisely, we verified by brute force that all non-trivial off-diagonal matrix elements of $h_{DD'}$ in Eq.~\eqref{eq:hDD} can be set equal to $-t$ in the first three sectors but not the last.] The frustration in the $(0, 0)$ sector has a similar origin as the honeycomb lattice model: flipping a collection of plaquettes that covers the entire torus yields a $\pi$ phase only in the $(0, 0)$ sector, while all other loops of dimer moves accumulate no net phase. 

We also performed exact diagonalizaton of our model on a $4 \times 4$ torus constrained to the restricted Hilbert space. At the RK point, we find one ground state in each of the $(1, 0), (0, 1), (1, 1)$ sectors with exactly zero energy. The perfectly staggered states of the triangular lattice dimer model remain zero-energy states at this point as well, but only three ground states extend into the region $V<t$, as in the bosonic dimer model. Assuming the unfrustrated condition persists all the way to the thermodynamic limit, the ground states of the $(1, 0), (0, 1), (1, 1)$ sectors are degenerate for the entire parameter region $V_c < V < t$ (where the bosonic dimer model is in the topological phase) when the vertex constraints are strictly enforced with $J_v, J_e \to \infty$. The lowest excited states in the exact diagonalization are the lowest-energy states in the $(0,0)$ sector---which for the $4 \times 4$ lattice at the RK point are $6$-fold degenerate with energy $0.14t$. By analogy with the Majorana loop model, these lowest excited states cost an energy comparable to inserting a vison into the bosonic dimer model, and thus we expect they remain gapped away from the other ground states in the thermodynamic limit. For the range of parameters where the ground state of the $(0, 0)$ sector has higher energy than the other sectors, the model will have three topologically degenerate fermion-parity-odd ground states on a torus as well as gapped edges with open boundary conditions. Thus it is natural to expect that the resulting topological order is identical to the above Majorana loop model.

\section{Identifying Topological Order} \label{sec:toporder}
We are now ready to analyze the universal properties of the gapped states obtained above. We will present both analytical and numerical evidence that the topological order indeed corresponds to an Ising phase together with a chiral $p_x-ip_y$ superconductor. For theoretical expedience we primarily concentrate on the Fisher-lattice model, which allows many exact statements to be made given the exact solvability. We stress, however, that the results are expected to extend straightforwardly to the triangular lattice as well. 

\subsection{Ising topological quantum field theory review} \label{app:Ising}
We first review the Ising topological quantum field theory (TQFT). This topological phase---which is realized, e.g., in Kitaev's honeycomb model~\cite{Kitaev06a} or the $\nu=1$ bosonic Pfaffian quantum Hall state~\cite{Greiter1991,Greiter1992}---supports three types of anyons denoted by $I, \sigma, \psi$.
The nontrivial fusion rules are given by
\begin{gather} \begin{split}
	\sigma\times\sigma&=I+\psi \\
	\sigma\times\psi&=\sigma \\
	\psi\times\psi&=I.
	\label{}
\end{split} \end{gather}
It turns out that eight different bosonic topological phases exhibit these same fusion rules yet are distinguished by the topological twist factor of $\sigma$:
\begin{align}
	\theta_\sigma&=e^{\frac{\pi in}{8}}, 	\label{}
\end{align}
where $n$ is an odd integer. (For any $n$ the $\psi$ twist factor is $\theta_\psi = -1$.) Since the corresponding chiral central charge is $c_-=n/2$, we label these phases as $\text{Ising}^{(n/2)}$. It is worth mentioning that the bulk-anyon braiding statistics is identical for $n$ and $n+16$. The usual Ising phase~\cite{Kitaev06a} is $\text{Ising}^{(1/2)}$ in this notation.

The modular matrices on a torus, which have been conjectured to uniquely
identify the topological phase~\cite{rowell2009,bruillard2013}, are given by~\cite{Kitaev06a}
\begin{align}
	\label{eq:Ising}
	S&=\frac{1}{2} \left(
\begin{array}{ccc}
 1 & 1 & \sqrt{2} \\
 1 & 1 & -\sqrt{2} \\
 \sqrt{2} & -\sqrt{2} & 0 \\
\end{array}
\right) \\
T&=e^{-\frac{\pi in}{24}}
\begin{pmatrix}
	1 & 0 & 0\\
	0 & -1 & 0\\
	0 & 0 & e^{\frac{\pi in}{8}}
\end{pmatrix}.
\end{align}
Here, $T$ encodes the self-statistics (twist factors) of the quasi-particles, while $S$ encodes the mutual statistics.

The ground-state degeneracy (GSD) of a topological phase on a genus-$g$ surface can be obtained from the Verlinde formula~\cite{Moore1990}:
\begin{equation}
	\text{GSD}=\sum_{a}S_{Ia}^{2-2g}
	\label{}
\end{equation}
with $a$ running over all quasiparticle types. For the Ising TQFT, we have $S_{II}=S_{I\psi}=\frac{1}{2}$ and $S_{I\sigma}=\frac{1}{\sqrt{2}}$, yielding
\begin{equation}
	\text{GSD}=2\cdot\Big(\frac{1}{2}\Big)^{2-2g}+\Big(\frac{1}{\sqrt{2}}\Big)^{2-2g}= 2^{g-1}(2^g+1).
	\label{eqn:IsingGSD}
\end{equation}

The systems under consideration arise microscopically from fermionic matter, so it is useful to consider Ising phases supplemented by physical fermions, whose particle content is denoted by $\text{Ising}^{(n/2)}\times \{I, f\}$. Now the self-statistics of an anyon is only defined up to $\pm 1$ since one can always attach a fermion $f$ to the anyon. Therefore, the bulk anyon properties are identical in this case for $n$ and $n\pm 8$. 

\subsection{Ground-State Degeneracy on Closed Surfaces}
The first piece of evidence that our Majorana-dimer models support Ising topological order comes from the ground-state degeneracy on closed surfaces. We have shown in Sec.~\ref{MajoranaModelsSec} that systems defined on a torus host a three-fold ground-state degeneracy. We now further argue that on a genus-$g$ surface the ground-state degeneracy is $2^{g-1}(2^g+1)$---exactly as for an Ising topological phase [see Eq. \eqref{eqn:IsingGSD}].

Recall that the three-fold degeneracy on a torus arises because only odd-fermion-parity states can maximally satisfy all Hamiltonian terms, implying that 
one of the ground states of the pure bosonic dimer model is lifted to higher energy.
Similar constraints hold on higher-genus surfaces. In fact, we will show that
\begin{equation}
	\prod_p \mathbf{B}_p^{\varhexagon} = (-1)^{g} P_f,
	\label{eqn:productBp2}
\end{equation}
where $g$ is the genus of the surface. 
This relation can be proven inductively. A surface $\Sigma_{g}$ with genus $g$ can be obtained from a genus-$(g-1)$ surface $\Sigma_{g-1}$ by making a hole and gluing on an open torus T. Without losing generality, we can choose a trivalent graph such that the gluing hole coincides with a plaquette $p_0$. Assuming the relation~\eqref{eqn:productBp2} holds for $\Sigma_{g-1}$, we have
\begin{equation}
	\mathbf{B}_{p_0}^{\varhexagon}\prod_{\substack{p\in \Sigma_{g-1}\\p\neq p_0}}\mathbf{B}_p^{\varhexagon} = (-1)^{g-1}P_f(\Sigma_{g-1}) .
\end{equation}
For the open torus (punctured at $p_0$), using Eq.~\eqref{eqn:productBp} we have
\begin{equation}
	\mathbf{B}_{p_0}^{\varhexagon}\prod_{\substack{p'\in \text{T}\\p'\neq p_0}}\mathbf{B}_{p'}^{\varhexagon}=-P_f(\text{T}).
	\label{}
\end{equation}
We then glue the open torus and $\Sigma_{g-1}$ together at $p_0$ to get $\Sigma_{g-1}$; after gluing the plaquette $p_0$ no longer belongs to the surface $\Sigma_g$.
Multiplying the two relations and recalling that $(\mathbf{B}_{p_0}^{\varhexagon})^2=1$ (in the restricted Hilbert space), we obtain
\begin{equation}
	\prod_{p\in \Sigma_g}\mathbf{B}_p^{\varhexagon} = (-1)^g P_f(\Sigma_{g-1})P_f(\text{T})=(-1)^g P_f(\Sigma_g),
	\label{}
\end{equation}
yielding Eq.~\eqref{eqn:productBp2} as claimed. Therefore, on a genus-$g$ surface all ground states must have fermion parity equal to $P_f=(-1)^g$ in order to satisfy $\mathbf{B}_p^{\varhexagon}=1~\forall~p$. Using the result of Appendix~\ref{app:Kasteleyn}, the number of ground state on a genus-$g$ surface is then $2^{g-1}(2^g+1)$.

The fact that the ground states have global fermion parity equal to $(-1)^g$ can be understood from the presence of the ``hidden'' $p_x-ip_y$ superconductor: it is known that the ground state of a $p_x-ip_y$ superconductor on a torus with periodic boundary conditions has odd fermion parity~\cite{Read00}. Generalizing to a genus-$g$ surface (which can be viewed as a connected sum of $g$ tori), it is not hard to see that the ground state fermion parity should be $(-1)^g$. Since the Ising phase is purely bosonic, the ground state fermion parity of the $\text{Ising}\times(p_x-ip_y)$ topological phase is also $(-1)^g$. 

\subsection{Fully Gapped Boundary to Vacuum}
For the discussion of the edge physics, it is important to fix the background in which the phase desribed here arises. In the following, we will take the Majorana degrees of freedom in our model to arise microscopically from a medium with zero ``background'' central charge, such as an array of Kitaev chains. As an alternative setup, the Majoranas could arise from a vortex lattice in a chiral $p$-wave superconductor with central charge $c = \pm 1/2$; results for the latter case can be obtained straightforwardly from the setup examined explicitly below.

Since we have shown that the Majorana loop state on the Fisher lattice is the ground state of a commuting-projector Hamiltonian, on a manifold with boundary there can not be any chiral edge modes~\cite{Kitaev06a}. We have also shown that the Majorana-dimer model on a triangular lattice is fully gapped with open boundary conditions. A fully gapped boundary implies the following: 
(a) The chiral central charge $c_-$ must vanish. For $\text{Ising}\times (p_x-ip_y)$ we indeed find that $c_- = \frac{1}{2}-\frac{1}{2}=0$. 
(b) The topological order must contain a ``Lagrangian subalgebra''~\cite{levin2013, kitaev2012, kong2014, eliens2013}, namely a set of bosonic quasiparticles whose condensation eliminates the topological order completely. For the $\text{Ising}\times(p_x-ip_y)$ topological phase, the particle content in the bulk can be conveniently represented by $\{I, \sigma, \psi\}\times \{I, f\}$ where $f$ represents physical fermions. One can easily identify the Lagrangian subgroup as $\{I, \psi f\}$. Condensing the combination $\psi f$ identifies $\psi$ with $f$ and confines both $\sigma$ and $\sigma f$ due to the nontrivial braiding statistics between $\sigma$ and $\psi$~\cite{bais2009}; the result is a trivial fermionic phase with particle content $\{I, f\}$. Together with the vanishing of the chiral central charge, this implies the existence of a fully gapped edge~\cite{Barkeshli2015}.

We notice that the $\psi f$ boson can be identified with the ``vison'' excitation of the lattice model. A vison in the Majorana loop model corresponds to a plaquette violation, i.e. $B_p=-1$ for a certain $p$. Such excitations can be generated with a string operator along an open path $P$ on the dual lattice:
\begin{equation}
	W_v(P)=\prod_{j\in P} \sigma^z_j.
	\label{}
\end{equation}
Here the product runs over all edges $j$ intersecting with $P$.
Notice that $W_v$ does not involve any Majoranas, and in fact takes the same form as the string operator that generates plaquette excitations in the bosonic toric code. Therefore we expect the visons are bosonic.

\subsection{Modular Matrices} \label{sec:modular}

The above arguments illustrate the consistency of the $\text{Ising}\times (p_x-ip_y)$ theory with the numerical observations thus far. However, we should notice that there are in fact four different types of topological order that are consistent with the ground-state degeneracy counting and existence of gapped boundaries. Following the notation laid out above, these correspond to $\text{Ising}^{(n/2)}\times (p_x-ip_y)^n$, where again $n$ is an odd integer and $\text{Ising}^{(1/2)}$ denotes the usual Ising phase. All such phases have $c_-=0$.  Moreover, we should regard $n$ and $n+8$ as representing the same phase~\cite{SET} since their bulk anyon content is identical [recall Sec.~\ref{app:Ising}].  Thus the four distinct states that we would like to discriminate amongst correspond to $n=1,3,5,7$. 

To affirmatively and unambiguously identify the topological order, we characterize the topological properties of the bulk anyons through the modular $S$ and $T$ matrices, which can be extracted using the entanglement properties of ground states on the torus~\cite{ZhangPRB2012}. This calculation is done for the Majorana-dimer model on the Fisher lattice (Sec.~\ref{sec:loop-model}), since the vanishing correlation length for the ground states negates the need to perform any finite-size scaling; for this reason, a minimal $2\times 2$ lattice on a torus suffices.
[For the triangular lattice model (Sec.~\ref{sec:dimer-model}), analogous calculations on small clusters (e.g., $4\times 4$) were inconclusive most likely due to the system's finite correlation length; we leave for future work a thorough numerical investigation of this model using, for example, DMRG.]
Because the ground states preserve the $C_3$ rotation symmetry of the Fisher lattice, we adopt the method developed in Refs.~\cite{ZhangPRB2012, CincioPRL2013, BauerSemion} to extract the modular matrix $ST^{-1}$ using the action of a $2\pi/3$ rotation. This allows us to compute $T$ and $S$ individually given minimal assumptions about the form of these matrices.  Without this rotational symmetry, we could instead use the methods of Ref.~\cite{ZhangPRB2015} to compute the $S$ matrix and constrain the $T$ matrix.

The presence of fermions in our model forces us to slightly modify the algorithm of Refs.~\cite{ZhangPRB2012, CincioPRL2013, BauerSemion} to determine the modular matrices $S$ and $T$. There are two assumptions of these previous works that no longer hold. The first is that the modular matrix $ST^{-1}$, which corresponds to a $2\pi/3$ rotation, satisfies $(ST^{-1})^3 = 1$, i.e., $R_{2\pi/3}^3=1$. Naively, one might define the rotation through its action on the the fermionic operator $f_q$ at site $q$ by $R_{2\pi/3}f_qR_{2\pi/3}^{-1}=f_{R_{2\pi/3}(q)}$. This would imply that the representation of $R_{2\pi/3}$ on the ground state manifold would have to satisfy $R_{2\pi/3}^3=1$. In a fermionic topological phase, however, rotations can act in more subtle ways.

To see this, we note that $S$ and $T$ matrices must be understood as the non-Abelian Berry phases associated with the degenerate ground states under adiabatic deformation of the system~\cite{Wen1993}, and thus we should view $R_{2\pi/3}$ in the same way for the purpose of extracting modular matrices.
As demonstrated in Ref.~\cite{YouCheng2015} via explicit Berry phase calculations, modular transformations of the ground state of a $(p_x-ip_y)^n$ superconductor with periodic boundary conditions along both directions are given by 
\begin{align}
	S_{(p_x-ip_y)^n}&=e^{\frac{\pi i n}{4}}
	&T_{(p_x-ip_y)^n}&=e^{-\frac{\pi i n}{12}}. 
	\label{eq:p-ip}
\end{align}
In particular, the ground state on a torus with $C_3$ symmetry satisfies
\begin{equation}
R_{\frac{2\pi}{3}}^3 = (ST^{-1})^3 = (e^{\frac{\pi i n}{3}})^3 = (-1)^n.
\end{equation}
Similarly, a ground state on a torus with $C_4$ symmetry satisfies $R_{\frac{\pi}{2}}^4 = S^4 = (e^{\frac{\pi i n}{4}})^4 = (-1)^n$.
The nontrivial right-hand side is a direct consequence of the fact that the ground state of a $(p_x - ip_y)^n$ superconductor has odd fermion parity when $n$ is odd, because a $2\pi$ rotation acting on a fermion yields a $-1$ phase factor---where again the $2\pi$ rotation should be understood in the sense of an adiabatic Berry phase. 
We will need to account for this subtle Berry phase effect in our calculation.
 
One can also obtain these relations by microscopic considerations in our setup: Adiabatically rotating the system by $2\pi$ can be seen to be topologically equivalent to a series of braids that for every Majorana operator sends $\gamma_i \rightarrow -\gamma_i$, and thus the action of $R_{2\pi/3}$ on the Majorana operators must be taken to be
\begin{equation}
  R_{2\pi/3} \gamma_i R_{2\pi/3}^{\dagger} = - \gamma_{R_{2\pi/3}(i)}. 
\end{equation}
Similar results can be obtained by viewing the system as a network of Majorana wires, where a $2\pi$ rotation is known to have the same effect~\cite{halperin2012}.

The second assumption that while valid for bosonic theories, must be reconsidered in our case, is that the modular matrix $S$ has a positive row and column corresponding to the vacuum anyon of the topological theory. While this assumption holds for bosonic topological orders, it can already be seen to fail for $S_{(p_x-ip_y)^n}$ above. The existence of a positive row and column can be used to extract $S$ and $T$ from a combination of modular matrices such as $ST^{-1}$~\cite{ZhangPRB2012}, but without it some ambiguity in the precise values of $S$ and $T$ persists. 
While these issues could be remedied by expensive adiabatic computations of the $S$ and $T$ matrices, we will show below that the easier minimally entangled state (MES) calculations indeed contain enough information to distinguish the  $\text{Ising}^{(n/2)}\times (p_x-ip_y)^n$ phases.
The key fact is that the modular matrices of these theories,  which read  
\begin{equation}
	\begin{gathered}
		S = S_{\text{Ising}^{(n/2)}} \otimes S_{(p_x-ip_y)^n}\\
		T =  T_{\text{Ising}^{(n/2)}} \otimes T_{(p_x-ip_y)^n}
	\end{gathered}
	\label{eq:ST}
,
\end{equation}
will still have a row and column that are positive \emph{modulo a constant prefactor}, since the difference from a bosonic theory is completely due to an overall phase contributed by the $(p_x-ip_y)^n$ sector. 
In the following, we will carefully step through the logic to show that the Majorana-dimer model of this paper produces the topological phase with $n=1$. 
Then we discuss how our construction can be modified to produce Hamiltonian and wavefunction representatives for each of the other odd $n$ as well.

To proceed we must first choose a basis $\{ \ket{i} \}$ for the three-dimensional ground-state manifold.  We employ the ground states $\ket{n_1,n_2}$ of the Majorana-dimer model in each topological sector with fixed winding numbers $n_1, n_2$ of the transition graph loops, where $(n_1, n_2)$ takes one of the three values $(0, 1), (1, 0), (1, 1)$.
Notice that the overall phase of each ground state is arbitrary and that the winding number basis does not clearly specify the phases.

The second step of the analysis is to compute the overlap matrix $\braket{i | R_{2\pi/3} | j}$ for these ground states. As discussed above, we choose
the rotation to act such that
\begin{equation}
R_{2\pi/3}^3 = P_f.
\end{equation}
One possible choice of phase convention is that the action of $2\pi/3$ rotations in the winding number basis takes the form
\begin{equation}
	R_{\frac{2\pi}{3}}=
	\begin{pmatrix}
	0 & 1 & 0\\
	0 & 0 & -1\\
	1 & 0 & 0
	\end{pmatrix},
	\label{eq:rotmatrix}
\end{equation}
which indeed satisfies $R_{2\pi/3}^3=-1$. Other phase conventions for the ground states yield a rotation matrix that differ from the above by conjugation with a diagonal matrix of phases, but do not affect the final answers below.

\begin{figure}
	  \begin{subfigure}
 	 \centering
 	 \includegraphics[angle=90, width=0.5\linewidth]{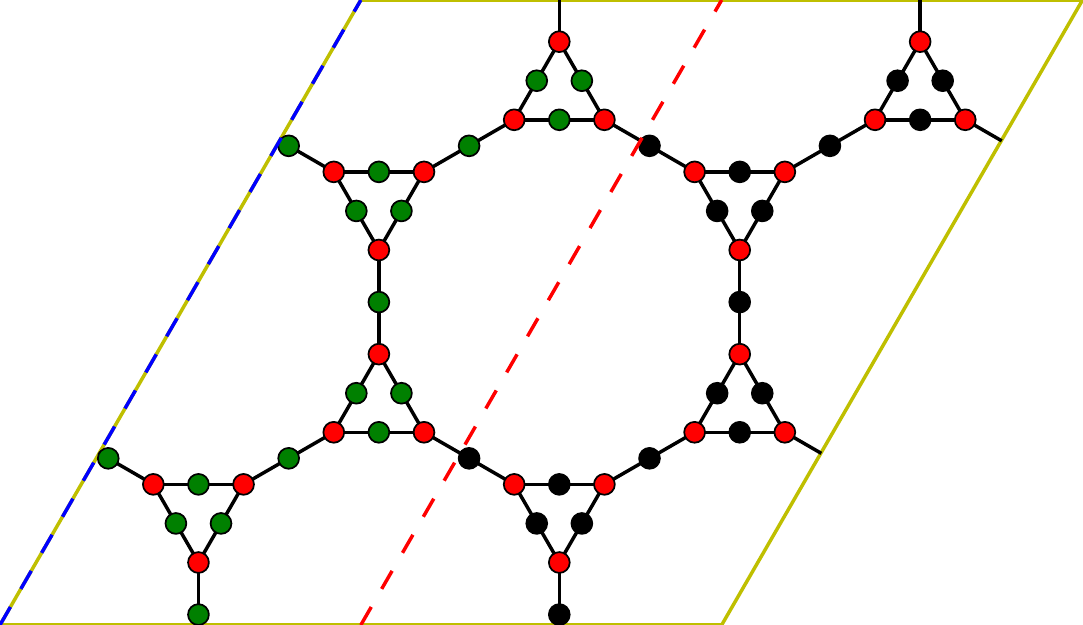}     
 	 \end{subfigure}
	\hspace{-1cm}
	\begin{subfigure}
    \centering
    \includegraphics[width=0.5\linewidth]{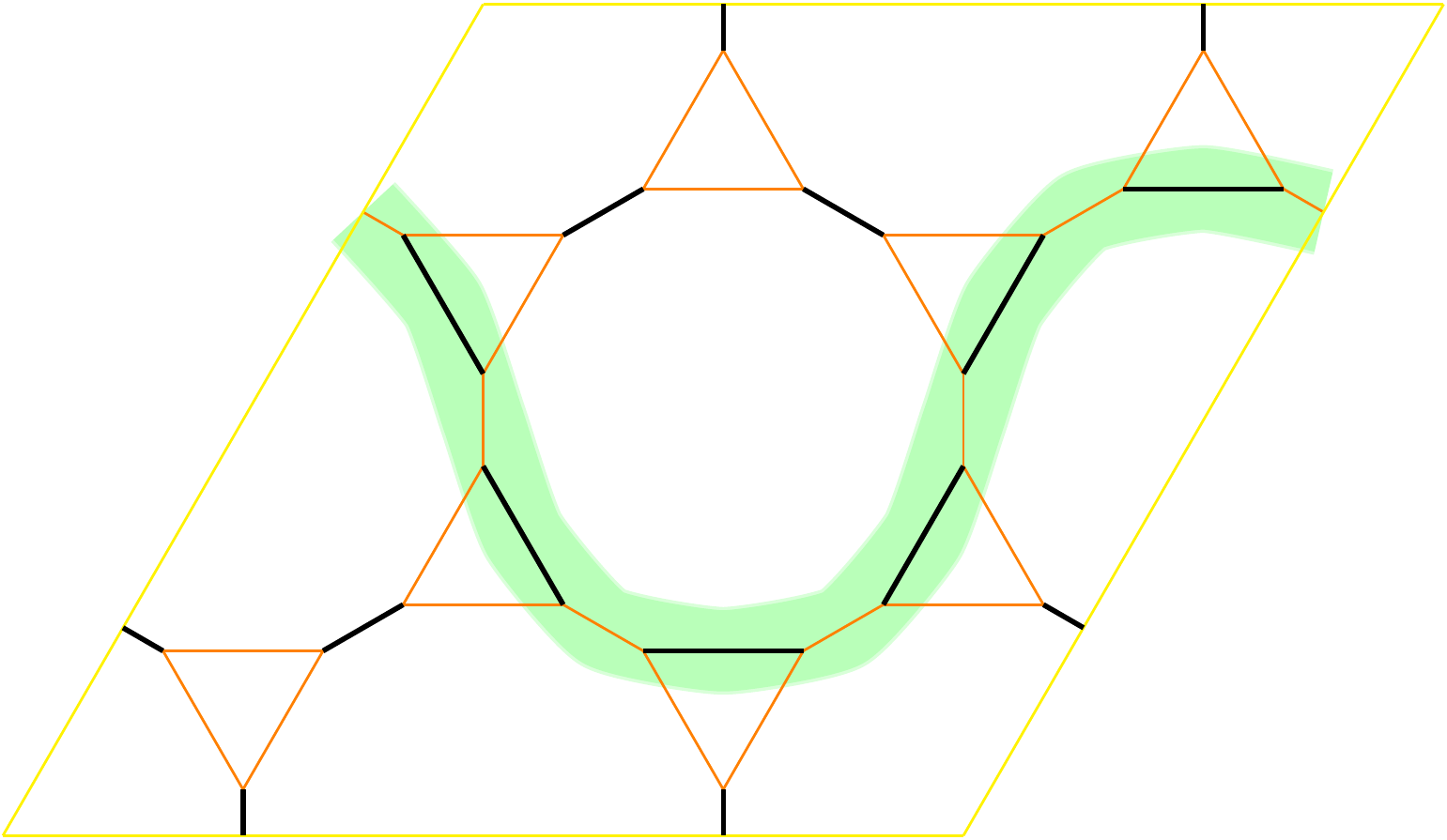} 
    \end{subfigure}
	\begin{subfigure}
		\centering
		\includegraphics[width=0.8\linewidth]{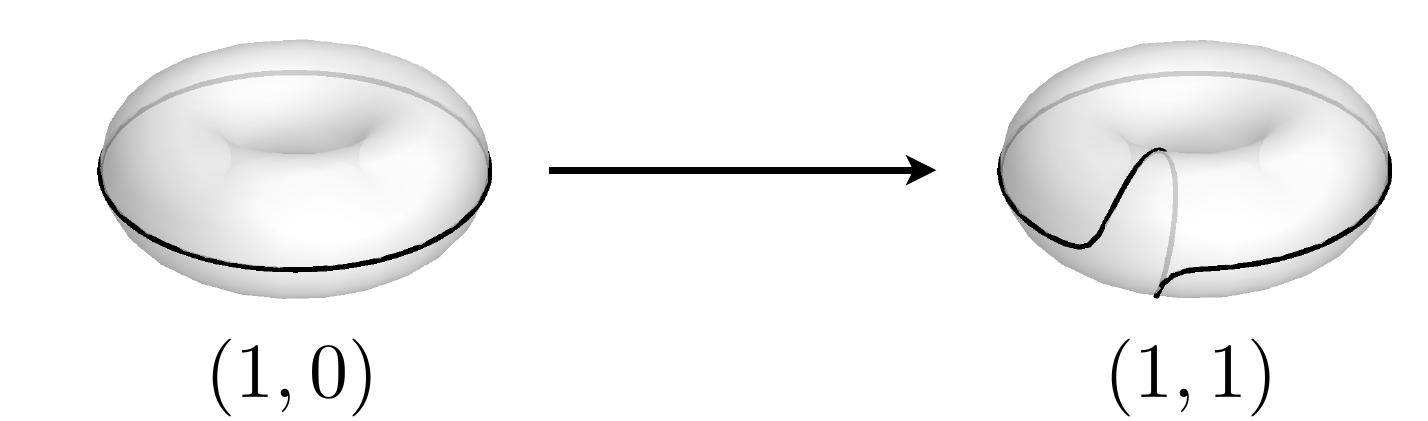}
	\end{subfigure}
  \caption{Upper left: The entanglement cut (red dashed line) used for the numerical calculations of the modular $S$ and $T$ matrices. Upper right: A dimer configuration belonging to the $(1, 0)$ topological sector. Bottom: The Dehn twist $T$ permutes the sectors $(1, 0)$ and $(1, 1)$ while preserving the sector $(0, 1)$. }
  \label{fig:cut}
\end{figure}

Accessing the anyon properties requires changing to the MES basis, i.e., the states that minimize the entanglement entropy with respect to a non-contractible cut of the torus~\cite{ZhangPRB2012}, which are known to have a definite topological charge through the torus. The MES basis for the cuts shown in Fig.~\ref{fig:cut} was found by brute-force minimization of the entanglement entropy. Using the phase convention defined in Eq.~\eqref{eq:rotmatrix}, we find
\begin{equation}
	\begin{split}
	\ket{1}&=\frac{1}{\sqrt{2}}(\ket{1,0}-e^{\frac{3i\pi}{8}}\ket{1,1})\\
	\ket{2}&=\frac{1}{\sqrt{2}}(\ket{1,0}+e^{\frac{3i\pi}{8}}\ket{1,1})\\
	\ket{3}&=\ket{0,1}.
	\end{split}
	\label{MES}
\end{equation}
The entanglement entropies of the three MES's are respectively $3\ln 2$, $3\ln 2$, and $4\ln 2$.  Generally speaking, a MES corresponding to anyon type $a$ should have topological entanglement entropy $\gamma_a=2\ln \frac{D}{d_a}$, where $d_a$ is the quantum dimension of $a$ and $D = \sqrt{\sum_a d_a^2}$ is the total quantum dimension~\cite{ZhangPRB2012}. For Ising anyons, we have $\gamma_I=\gamma_\psi=2\ln 2, \gamma_\sigma=\ln 2$. Up to topological-sector-independent area law contributions, this is fully consistent with the calculated entanglement entropies if we identify $\ket{3}$ with the non-Abelian $\sigma$ anyon.  

In the MES basis, the $2\pi/3$ rotation becomes
\begin{equation}
	R_{\frac{2\pi}{3}}=
	e^{\frac{3\pi i}{8}}
	\begin{pmatrix}
		\frac{1}{2} & -\frac{1}{2} & \frac{e^{\frac{\pi i}{4}}}{\sqrt{2}}\\
		\frac{1}{2} & -\frac{1}{2} & -\frac{e^{\frac{\pi i}{4}}}{\sqrt{2}}\\
		\frac{e^{-\frac{3\pi i}{8}}}{\sqrt{2}} & \frac{e^{-\frac{3\pi i}{8}}}{\sqrt{2}} & 0
	\end{pmatrix}.
	\label{eq:rotmatrix_mes}
\end{equation}
Following~\cite{CincioPRL2013} and~\cite{BauerSemion}, in a topologically ordered phase a $2\pi/3$ rotation of a torus is represented by $ST^{-1}$, up to conjugation by a diagonal phase matrix $D$ and a permutation matrix $P$: 
\begin{equation}
	R_{\frac{2\pi}{3}}= P D \left( ST^{-1} \right) D^\dag  P^\dag.
	\label{eq:rotation}
\end{equation}
The undetermined matrices $D$ and $P$ are due to the freedom to rephase each MES and reorder the MES's with the same topological entanglement entropy. 
Here both $T$ and $D$ are diagonal, while $S$ is proportional to a matrix with all positive elements in the first row and column. We will consider this equation for each possible permutation $P$.

The above equation with $P=I$ allows us to determine $S$ up to an overall phase by fixing the first row and column of $R_{\frac{2\pi}{3}}$ to be non-negative as in Eq.~\eqref{eq:rotmatrix}:
\begin{equation}
	S=
	e^{i \eta}
	\begin{pmatrix}
		\frac{1}{2} & \frac{1}{2} & \frac{1}{\sqrt{2}}\\
		\frac{1}{2} & \frac{1}{2} & -\frac{1}{\sqrt{2}}\\
		\frac{1}{\sqrt{2}} & -\frac{1}{\sqrt{2}} & 0
	\end{pmatrix}.
	\label{eq:S}
\end{equation}
This is, not surprisingly, the $S$ matrix of an Ising topological phase up to an overall phase.
By matching this to Eq.~\eqref{eq:rotation}, we can solve for $T$:
\comment{
\begin{equation*}
	D = e^{i (\frac{3\pi}{8} - \eta)}	
	\begin{pmatrix}
		1 & 0 & 0\\
		0 & 1 & 0\\
		0 & 0 & e^{-\frac{3\pi i}{8}}
	\end{pmatrix} , \, T^{-1}D^\dag = 
    \begin{pmatrix}
		1 & 0 & 0\\
		0 & -1 & 0\\
		0 & 0 & e^{\frac{\pi i}{4}}
	\end{pmatrix}
\end{equation*}
and thus
}
\begin{equation}
 T = e^{i (\eta - \frac{3\pi}{8})}
 \begin{pmatrix}
		1 & 0 & 0\\
		0 & -1 & 0\\
		0 & 0 & e^{\frac{\pi i}{8}}
	\end{pmatrix}.	
	\label{eq:T}
\end{equation}
This form of $T$ and the knowledge that $c_- = 0$ pins down the topological order to be $\text{Ising}^{(1/2)}\times (p_x-ip_y)$. 
In particular, upon selecting $\eta = \frac{\pi}{4}$ these $S$ and $T$ matrices precisely agree with Eq.~\eqref{eq:ST} with $n=1$
\footnote{
More generally, selecting $\eta = \frac{\pi x}{4}$, one can check that the $S$ and $T$ matrices will agree with those of $\text{Ising}^{(1/2)}\times (p_x-ip_y)^x$ whenever $x=1 \mod 6$. This is due to the fact that $6$ copies of $(p_x-ip_y)$ contributes a factor of $1$ to the $ST^{-1}$ matrix --- thus, the modular matrix $ST^{-1}$ does not completely determine the topological phase. Since $6$ copies of $(p_x-ip_y)$ shift the chiral central charge $c_-$ by $3$, the additional knowledge that $c_- = 0$ fixes the appropriate factor $(p_x-ip_y)^x$.}.

Suppose that we instead choose a permutation matrix $P$ that swaps the states $|1\rangle$ and $|2\rangle$ that possess identical topological entanglement entropies.  While the same $S$ results, we find that the $T$ matrix now takes the form
\begin{equation}
 T = e^{i (\eta + \frac{5\pi}{8})}
 \begin{pmatrix}
		1 & 0 & 0\\
		0 & -1 & 0\\
		0 & 0 & e^{\frac{ 9 \pi i}{8}}
	\end{pmatrix}.	
	\label{eq:T9}
\end{equation}
Now the $S$ and $T$ matrices agree with Eq.~\eqref{eq:ST} with $n=9$ when $\eta = \frac{9\pi}{4}$. As discussed above, the phases $n=1$ and $n=9$ have the same bulk anyon content and should be identified; thus, this result is consistent with a unique identification of the topological phase from the modular matrices and chiral central charge.

\subsubsection{Semion Variant}\label{sec:semion}

A similar calculation can be performed with the semion version of the dimer model introduced in Sec.~\ref{sec:loop-model}. In this case, we instead find the MES to be given by 
\begin{equation}
	\begin{split}
	\ket{1}&=\frac{1}{\sqrt{2}}(\ket{0,1}+e^{\frac{7i\pi}{8}}\ket{1,1})\\
	\ket{2}&=\frac{1}{\sqrt{2}}(\ket{0,1}-e^{\frac{7i\pi}{8}}\ket{1,1})\\
	\ket{3}&=\ket{1,0}.
	\end{split}
	\label{eq:semionMES}
\end{equation}
Using this MES basis, the rotation matrix is written as 
\begin{equation}
	R_{\frac{2\pi}{3}}=
	e^{-\frac{\pi i}{8}}
	\begin{pmatrix}
		\frac{1}{2} & -\frac{1}{2} & -\frac{e^{\frac{\pi i}{4}}}{\sqrt{2}}\\
		\frac{1}{2} & -\frac{1}{2} & \frac{e^{\frac{\pi i}{4}}}{\sqrt{2}}\\
		\frac{e^{\frac{\pi i}{8}}}{\sqrt{2}} & \frac{e^{\frac{\pi i}{8}}}{\sqrt{2}} & 0
	\end{pmatrix}.
\end{equation}
This leads to the same $S$ matrix as Eq.~\eqref{eq:S}, but a distinct $T$ matrix:
\begin{equation}
 T = e^{i (\eta + \frac{\pi}{8})}
 \begin{pmatrix}
		1 & 0 & 0\\
		0 & -1 & 0\\
		0 & 0 & e^{\frac{5\pi i}{8}}
	\end{pmatrix}.	
\end{equation}
With $\eta = \frac{5\pi}{4}$, the $S$ and $T$ now match the product of an $\text{Ising}^{(5/2)}$ theory and a $(p_x - i p_y)^5$ superconductor. Intuitively, the topological twist of $\sigma$ shifts by $i$ due to the attachment of a semion with exchange statistics $i$ to the Ising anyon.

\section{Generalizations to multiple Majoranas per site: 8-fold way}\label{sec:8fold}

\begin{table}
	\centering
\begin{tabular}{ l c l }
  $n$ & Phase & Twists  \\
  \midrule
  $0$ & Toric Code & $1, -1, 1, 1$ \\
  $2$ & $\mathrm{U}(1)_4$ & $1, -1, e^{i\pi/4}, e^{i\pi/4}$\\
  $4$ & $\mathrm{U}(1)_2\times\mathrm{U}(1)_2$ & $1, -1, e^{i\pi/2}, e^{i\pi/2}$\\
  $6$ & $\mathrm{SO}(6)_1$ & $1, -1, e^{3i\pi/4}, e^{3i\pi/4}$\\
  $8$ & $\mathrm{SO}(8)_1$ & $1, -1, -1, -1$ \\
\end{tabular}
\caption{Even-$n$ phases from the 16-fold way. \label{tab:evenn} }
\end{table}

The eight topological phases $\text{Ising}^{(n/2)}$ for $n = 1, 3, 5, \ldots, 15$ discussed in Sec.~\ref{app:Ising} can be generated using a procedure of tensoring and condensation of bosons~\cite{bais2009, Neupert2016-bs}. Specifically, tensoring $n$ layers of Ising topological phases and condensing all of the bosons $\psi_i \psi_{i+1}$ formed from the combined fermions of neighboring layers gives the above progression of phases for odd $n$.
For even $n \leq 8$, the phases listed in Table~\ref{tab:evenn} occur.
The phases for $8<n<16$ can be described as conjugates of the $n' = 16-n$ phases listed in the table. The phase $n=16$ has identical bulk particle content as Kitaev's toric code, and the pattern repeats with period $16$. This is Kitaev's 16-fold way for gauged topological superconductors~\cite{Kitaev06a}.
A similar progression of phases occurs using the fermionic $\text{Ising}^{(1/2)}\times (p_x-ip_y)$ of this paper as a generating state. However, as mentioned in Sec.~\ref{app:Ising}, the pattern repeats after $n=8$, since the topological twists are only well defined up to an overall sign in the presence of physical fermions. 

To create Majorana-dimer wavefunctions for these $n>1$ phases, it suffices to accompany each bosonic dimer with $n$ Majorana-dimers instead of just $1$. Specifically, each lattice site $i$ now has $n$ Majorana modes $\gamma_i^{(\alpha)}, \alpha \in \{1, \ldots, n\}$, and the $n$ Majorana-dimers $(\gamma_i^{(\alpha)},\gamma_j^{(\alpha)})$ are formed with the same orientation for each $\alpha$. These wavefunctions can be viewed as dressed loop wavefunctions with $n$ copies of the Kitaev chain along each loop. It is straightforward to write down an exactly solvable Hamiltonian for this state by generalizing the construction in Sec.~\ref{sec:loop-model}.

To see that this procedure suffices, imagine $n$ initially decoupled copies of the $n=1$ model. Now add a coupling term between neighboring layers that energetically favors the loops in each layer to reside at the same location. Since the vison is just the violation of the plaquette term, and the simultaneous violation of plaquettes in two layers is invisible if loops are forced to surround the same plaquette in each layer, each of these terms drives a condensation transition that condenses the vison pairs $\psi_i\psi_{i+1}$ from neighboring layers (recall that the vison in each of the layer is the $\psi_i f$ particle). The end result of this process is the same as a single layer of loops dressed by $n$ copies of the Kitaev chain.

We bolster the above argument by repeating the calculation of the modular matrices for the $n=2$ case. Here we have four ground states---all with even fermion parity---that are formed from the four topological sectors. Using a phase convention where the rotation matrix takes the form
\begin{equation}
	R_{\frac{2\pi}{3}}=
	\begin{pmatrix}
    1 & 0 & 0 & 0\\	
	0 & 0 & 1 & 0\\
	0 & 0 & 0 & 1\\
	0 & 1 & 0 & 0
	\end{pmatrix},
\end{equation}
the MES were found to be 
\begin{equation}
	\begin{split}
	\ket{1}&=\frac{1}{\sqrt{2}}(\ket{1,0}-e^{\frac{i\pi}{4}}\ket{1,1})\\
	\ket{2}&=\frac{1}{\sqrt{2}}(\ket{1,0}+e^{\frac{i\pi}{4}}\ket{1,1})\\
	\ket{3}&=\frac{1}{\sqrt{2}}(\ket{0,0}+\ket{0,1})\\
	\ket{4}&=\frac{1}{\sqrt{2}}(\ket{0,0}-\ket{0,1}).
	\end{split}
\end{equation}
Using the same procedure as before, we find that the modular matrices satisfy
\begin{equation}
	S=
	\frac{e^{i \eta}}{2}
	\begin{pmatrix}
		1 & 1 & 1 & 1\\		
		1 & 1 & -1 & -1\\
		1 & -1 & -i & i\\
		1 & -1 & i & -i
	\end{pmatrix},
\end{equation}
and 
\begin{equation}
 T = e^{i (\eta - \frac{3\pi}{4})}
 \begin{pmatrix}
		1 & 0 & 0 & 0\\
		0 & -1 & 0 & 0\\
		0 & 0 & e^{\frac{\pi i}{4}} & 0 \\
        0 & 0 & 0 & e^{\frac{\pi i}{4}}	
	\end{pmatrix}.	
\end{equation}

These are precisely the modular matrices of the $\mathrm{U}(1)_4\times (p_x-ip_y)^2$ theory with $\eta = \frac{\pi}{2}$. As in Section~\ref{sec:modular}, permutations of the MES produce different forms for the $T$ matrix with the same $S$ matrix\comment{\footnote{Up to overall minus signs in the twists, the only other $T$ matrix appears that is not obviously equivalent to the one given is
\begin{equation*}
 T = e^{i \eta}
 \begin{pmatrix}
		1 & 0 & 0 & 0\\
		0 & 1 & 0 & 0\\
		0 & 0 & e^{-\frac{\pi i}{4}} & 0 \\
        0 & 0 & 0 & e^{\frac{3 \pi i}{4}}	
	\end{pmatrix}. 
\end{equation*} It can be shown that this $T$ matrix results by ?.}}, but these can all be regarded as representing the same bulk topological order. We note that a different exactly solvable model for this fermionic topological phase was studied in \Ref{GuPRB2014b}. 

In the above construction, each additional layer and condensation of $\text{Ising}^{(1/2)}\times (p_x-ip_y)$ increases $n$ by $1$. We can also \emph{decrease} $n$ by using a layer of a conjugate phase. One way to produce the conjugate phase is to act on the $n=1$ state with an anti-unitary operator $T$, such as 
\begin{gather}
T i T^{-1} = -i \\
T \gamma_i T^{-1} = \gamma_i.
\end{gather}
This operation flips the sign of the coupling of all Majorana pairs, and so is equivalent to reversing the orientation of all Majorana-dimers. A repeat of the modular matrix calculation confirms that this produces the $n=-1$ state, or equivalently the $n=7$ state. Similarly, reversing the orientation of the $n = 5$ semion variant in Section~\ref{sec:semion} produces the $n=3$ state. Thus our single-layer states and their conjugates suffice to generate all four of the $n$-odd topological phases.  

One final check on the arguments in this section is provided by considering a layer construction of the $n=1$ and $n=-1$ states. This is done similarly to the $n=2$ construction, but with the orientation on one layer reversed. The modular matrix computation for this state produces the $S$ and $T$ matrix of the toric code topological order, which is the $n=0$ phase of the $8$-fold way.

\section{Conclusion} \label{sec:conclusions}
In this paper, we have introduced a new class of models, termed \emph{Majorana-dimer} models. Starting from models of bosonic dimers, we introduce Majorana modes on the edges of the lattice and couple them to the dimers such that the Majorana modes always pair up according to the dimer configurations. We explicitly construct two frustration-free Hamiltonians governing the dynamics of the dimers: an exactly solvable Hamiltonian consisting of commuting projectors on the Fisher lattice, and a much simpler Hamiltonian on the triangular lattice. We characterize the universal topological properties of the models using ground-state degeneracy on closed surfaces and modular transformations, and show that the resulting gapped phases realize $\text{Ising}\times (p_x-ip_y)$ topological order in the simplest case, and phases related to Kitaev's 16-fold way in the general case. All these phases have gapped boundaries and cannot arise in purely bosonic systems. We note that similar results have been obtained by Walker and Wang~\cite{Walker_unpub}. It is interesting to ask whether the phase described here is part of an even larger family of systems. A natural extension of our work would be to replace the Majorana modes by parafermionic generalizations~\cite{FendleyParafermions,barkeshli2012a, clarke2013, cheng2012, lindner2012,ParafermionReview}, and couple them to dimers to form a phase with deconfined excitations that harbor parafermion zero modes. 

The models we study can be viewed as gauged fermionic SPTs protected by an on-site $\mathbb{Z}_2$ symmetry~\cite{QiNJP2012, RyuPRB2012, HongPRB2013, GuPRB2014}. This is particularly clear in the Fisher lattice model: performing a duality transformation sends $\sigma^z_e\leftrightarrow  \tau_p^z\tau_q^z$, where $p$ and $q$ label the two plaquettes adjacent to $e$ and $\tau$'s are Ising spins on the dual lattice. This dual model has a global $\mathbb{Z}_2$ symmetry generated by $\prod_p \tau_p^x$, and loops in the original model correspond to Ising domain walls in the dual model. A commuting-projector model for this fermion SPT was recently found in \Ref{Tarantino2016-dn}, which is closely related to the Fisher lattice model studied in this paper via the above duality transformation. Moreover, the generalization to $n$ Majoranas per site discussed in Sec.~\ref{sec:8fold} can also be dualized to capture other $\mathbb{Z}_2$ fermionic SPT's, and the $8$-fold way precisely corresponds to the $\mathbb{Z}_8$ classification of the SPT's~\cite{RyuPRB2012, GuPRB2014}. These phases can also be realized with non-interacting fermions: consider the $n=1$ case and spin-$1/2$ electrons. Spin-up (down) electrons form $p_x+ip_y$ ($p_x-ip_y$) superconductors. The $\mathbb{Z}_2$ symmetry is generated by $(-1)^{N_\uparrow}$, i.e., conservation of the fermion parity of spin-up electrons. Gauging the $\mathbb{Z}_2$ symmetry would turn the $p_x+ip_y$ superconductor into an Ising phase, and therefore the gauged SPT is indeed $\text{Ising}\times (p_x-ip_y)$. 

Our results have important consequences for the question of which topological phases of matter can be represented with tensor
network states of small bond dimension. Previously, it has been shown that all bosonic topological phases with fully gapped boundaries
have exact PEPS representations~\cite{BuerschaperPRB2009, GuPRB2009, Bultinck_unpub}. At the same time, there is evidence
that topological phases with chiral edges and exponentially decaying bulk correlations---including the Ising theory whose particle
content is the same as the phase described here---cannot be efficiently represented as tensor networks~\cite{wahl2013,dubail2015,yang2015}. 
Crucially, given that we have explicitly constructed frustration-free Hamiltonians, the phase of matter discussed in this paper is likely to be
described exactly by a PEPS of relatively small bond dimension. Our construction therefore suggests that the use of fermionic systems
allows a broader class of topological orders to be desribed as tensor networks than previously known.
These phases may also be more susceptible to many-body localization~\cite{huse2013,bauer2013,potter2015,potter2016}.

Finally, it would be very interesting to realize the Majorana-RVB physics encapsulated in Eq.~\eqref{eq:RKWF}, and the resulting $\text{Ising}\times(p_x-ip_y)$-type topological order, in a purely fermionic microscopic setting (\emph{without} the accompanying bosonic dimers). In this context, our results highlight the possibility of a topological superconductor with $p$-wave pairing that breaks time reversal symmetry, but nevertheless has a gapped edge. This could be consistent with phenomenology observed in strontium ruthenates~\cite{maeno2001}.
Barring the admittedly far-fetched possibility of relevance to this material, one could obtain more natural models for this phase in engineered quantum systems. As a concrete physical realization we imagine, for example, a triangular Abrikosov vortex lattice in a two-dimensional $p_x+ip_y$ superconductor where each vortex hosts a Majorana zero mode~\cite{Read00} (for possible physics arising in such systems, see e.g., Refs.~\cite{grosfeld2006,ludwig2011,lahtinen2012,Franz15_PRB_91_165402}), or an appropriately arranged array of Majorana nanowires~\cite{kitaev2001,1DwiresLutchyn,1DwiresOreg, Barkeshli_unpub2015}. In each case, we at least have the correct Majorana degrees of freedom at hand. The question of whether one can design suitable interactions among these zero modes to induce an $\text{Ising}\times(p_x-ip_y)$-type phase must be addressed in future work, but the results presented here provide new motivation to address this problem.

\acknowledgments
We gratefully acknowledge Chetan Nayak for explaining the results in Refs.~\cite{freedman2011,freedman2011b}; Dave Aasen, Xie Chen, Zheng-Cheng Gu, Roman Lutchyn and Hong-Hao Tu for helpful discussions; and Kevin Walker for explaining his unpublished work.
This work was supported by the NSF through grant DMR-1341822 (JA and JHS); the Caltech Institute for Quantum Information and Matter, an NSF Physics Frontiers Center with support of the Gordon and Betty Moore
Foundation; and the Walter Burke Institute for Theoretical Physics at Caltech.
\appendix

\section{Fermion Parity Details: Clockwise-Odd Rule and State Counting} \label{app:Kasteleyn}

Let us prove the clockwise-odd rule for fermion parity on a transition loop. Consider a transition loop containing $2N$ Majoranas. Assume that the reference dimer configuration corresponds to the pairings $is_{2j-1,2j}\gamma_{2j-1}\gamma_{2j}=1$ with $j=1, \dots, N$, and that the new configuration has $is_{2j,2j+1}\gamma_{2j}\gamma_{2j+1}=1$. Denoting the loop fermion parity operator by $\hat P_{\rm loop}$, the fermion parity of the new state $\ket{\Psi}$ is \begin{equation}
	\begin{split}
		&\langle \Psi| \hat P_{\rm loop}|\Psi \rangle = \braket{\Psi|\prod_{j=1}^N is_{2j-1,2j}\gamma_{2j-1}\gamma_{2j}|\Psi}\\
	=&i^N\prod_{j=1}^N s_{2j-1,2j} \braket{\Psi|\gamma_1\gamma_2\cdots\gamma_{2N-1}\gamma_{2N}|\Psi}\\
=&-i^N\prod_{j=1}^N s_{2j-1,2j} \braket{\Psi|\gamma_2\cdots\gamma_{2N-1}\gamma_{2N}\gamma_1|\Psi}\\
=&-\prod_{j=1}^{2N} s_{j, j+1} \braket{\Psi|\prod_{j=1}^N is_{2j, 2j+1}\gamma_{2j}\gamma_{2j+1}|\Psi}\\
=&-\prod_{j=1}^{2N} s_{j, j+1}.
	\end{split}
		\label{}
\end{equation}
This is exactly the clockwise-odd rule quoted in Sec.~\ref{sec:parity}. 

We now consider the fermion parity of Majorana-dimer states on a high-genus surface, assuming periodic boundary conditions. There are $2^{2g}$ topological sectors of dimer configurations. Notice that a genus-$g$ surface can be viewed as the connected sum of $g$ tori. Each torus inherits the periodic boundary conditions so there are three states with odd fermion parity and one with even fermion parity. Therefore, the total number of states with even fermion parity is
\begin{equation}
	\begin{split}
	\sum_{k=0}^{k\leq[g/2]}\binom{g}{2k}3^{2k}&=\frac{1}{2}[(3+1)^g + (-1)^g (3-1)^g] \\
	&= \frac{1}{2}[2^{2g} + (-1)^g 2^{g}]\\
	&= 2^{g-1}[2^g + (-1)^g],
	\end{split}
\end{equation}
and the total number of states with odd fermion parity is $2^{2g}-2^{g-1}[2^g + (-1)^g]=2^{g-1}[2^g-(-1)^g]$.

\section{Majorana-Dimer Model with Open Boundary Conditions}
\label{sec:obc}

In Sec.~\ref{sec:hamiltonian} we defined a map from the Majorana-dimer model in the restricted Hilbert space to the bosonic dimer Hamiltonian via the nonlocal transformation $\ket{F(D)}\ket{D}\rightarrow \ket{D}$.  Recall that matrix elements for the fermionic part of the flip term are given by
\begin{equation}
	h_{DD'}=-\braket{F(D')|\mathcal{B}_p|F(D)}.
	\label{hApp}
\end{equation}
We fix the innate phase ambiguity for $\ket{F(D)}$ by the following convention.  Define $\ket{\mathbf{0}}$ as the vacuum of fermions in the reference configuration. The overlap $\braket{\mathbf{0}|F(D)}$ is always non-zero with open boundary conditions:
if we examine the transition graph between $D$ and the reference dimer configuration, $\ket{F(D)}$ is essentially the ground state of Kitaev chains on the transition loops, or in other words the state obtained by applying $\prod\limits_{e=(i,j) \in l} \frac{1 +is_{ij} \gamma_i \gamma_j}{{2}} $ along each transition loop $l$.  In our system with open boundary conditions, the ground state of each Kitaev chain has even fermion parity, and it is a well-known fact that the wavefunction of such chains is an equal-weight superposition (up to signs) of all fermion occupation numbers with given parity, including the vacuum.  
Thus with open boundary conditions $\braket{\mathbf{0}|F(D)}$ is indeed always non-zero, and we select phase conventions for $|F(D)\rangle$ such that this overlap is real and positive.  
 
Consider now the action of triangular-lattice flip term as defined in Eq.~\eqref{eqn:dimer-flip} for a certain plaquette $p$, and let the dimer configurations before and after the dimer flip by $D$ and $D'$, respectively.  (We assume that $D$ is flippable.)  Denote the fermionic part of $\mathbf{B}_p$ by 
\begin{equation}
  \mathcal{B}_p=e^{i s_{1p} \theta_{p}}\left(\frac{1 + s_{2p}\gamma_{p,1}\gamma_{p,2}}{\sqrt{2}}\right),
  \label{Bp}
\end{equation}
where $\gamma_{p,1}$ and $\gamma_{p,2}$ sit opposite the interior bond of the plaquette and $s_{1,2p}$ are signs that depend on the specific plaquette flip under consideration.  
[This expression is somewhat schematic but all we need here; see Eq.~\eqref{eqn:dimer-flip} for the precise form].  The main objective of this Appendix is to prove that with open boundary conditions
\begin{equation}
	\ket{F(D')} = \mathcal{B}_p \ket{F(D)}.
	\label{BpResult}
\end{equation}
Clearly $\ket{F(D')}$ and $\mathcal{B}_p \ket{F(D)}$ can at most differ by a phase factor, as they are both normalized and correspond to the same pairings of Majoranas. So to prove the equality it suffices to show that $\bra{\mathbf{0}}\mathcal{B}_{p}\ket{F(D)}>0$.

One can prove this relation by examining the transition graph between the configuration $D$ and the reference dimer configuration. There are three possible situations: 

 \begin{enumerate}
	 \item
 The first case arises when two dimers in the plaquette $p$ of the configuration $D$ belong to different transition loops, e.g., 
\begin{equation*}
 \begin{tikzpicture}[baseline={($ (current bounding box) - (0,0) $)}, >=stealth, scale=0.7]
	\draw [fill = black](0,0) circle (3pt);
	\draw [fill = black](1,0) circle (3pt);
	\draw [fill = black](0.5,0.87) circle (3pt);
	\draw [fill = black](1.5,0.87) circle (3pt);

    \draw[ultra thick] (0,0) -- (1,0);
    \draw[ultra thick] (0.5,0.87) -- (1.5,0.87);
    \draw[densely dotted] (0,0) -- (0.5,0.87);
    \draw[densely dotted] (1,0) -- (1.5,0.87);
    \draw[densely dotted] (0.5,0.87) -- (1,0);
    \draw[thick] (0.5,0.87) .. controls (-1.5,1.37) and (3.5,1.37) .. (1.5,0.87);
    \draw[thick] (0,0) .. controls (-2,-0.5) and (3,-0.5) .. (1,0);  
\end{tikzpicture},
\end{equation*}
and flipping them decreases the total number of loops in the graph by one.
Recall that the fermionic wavefunction of each transition loop in $D$ can be viewed as a Kitaev chain, and that the Kastelyn orientation guarantees that the fermion parity of the wavefunction (only counting those Majoranas on the transition loop) must be even.  Applying a Majorana operator $\gamma_i$ to a given loop flips the loop's parity and yields a wavefunction that is orthogonal to any wavefunction where that loop has even parity---including the reference state. Therefore we conclude that $\braket{\mathbf{0}|\gamma_{p,1}\gamma_{p,2}|F(D)}=0$ in this case. We should also notice that $e^{i\theta_p}=1$.  In our model the phase $e^{i \theta_p}$ is nontrivial only when the interior bond in the flipped plaquette coincides with a reference dimer.  But if that is the case then the two dimers that we flip must initially belong to the same transition loop, contradicting our assumption. We thus conclude from Eq.~\eqref{Bp} that 
\begin{equation}
	\bra{\mathbf{0}}\mathcal{B}_{p}\ket{F(D)}=\frac{1}{\sqrt{2}}\braket{\mathbf{0}|F(D)}>0.
	\label{case1}
\end{equation}

\item
When the two dimers in the plaquette $p$ originate from the same transition loop, the associated Kasteleyn arrows become important.  From Fig.~\ref{fig:orientation} we see that the arrows on the two dimers can orient either parallel or antiparallel.  The second case we consider arises when these arrows are antiparallel, e.g., 
\begin{equation*}
 \begin{tikzpicture}[baseline={($ (current bounding box) - (0,0) $)}, >=stealth, scale=0.7]
    \draw [fill = black](0,0) circle (3pt);
	\draw [fill = black](1,0) circle (3pt);
	\draw [fill = black](0.5,0.87) circle (3pt);
	\draw [fill = black](1.5,0.87) circle (3pt);

    \draw[->-,ultra thick] (0,0) -- (1,0);
    \draw[-<-,ultra thick] (0.5,0.87) -- (1.5,0.87);
    \draw[densely dotted] (0,0) -- (0.5,0.87);
	\draw[densely dotted] (1,0) -- (1.5,0.87);
	\draw[densely dotted] (1,0) -- (0.5,0.87);
    \draw[->-,thick] (1,0) .. controls (2,0) and (2.5,0.87) .. (1.5,0.87); 
    \draw[->-,thick] (0.5,0.87) .. controls (-0.5,0.87) and (-1,0) .. (0,0);
\end{tikzpicture}
\end{equation*}
Now the plaquette flip increases the number of loops by one.  This is exactly the inverse process of flipping two dimers belonging to different loops, so we can simply adopt the argument in case $1$ above to arrive at the same conclusion in Eq.~\eqref{case1}.

 \item
 The third case arises when two dimers with parallel Kasteleyn arrows belong to the same loop, e.g.,
 \begin{equation*}
 \begin{tikzpicture}[baseline={($ (current bounding box) - (0,0) $)}, >=stealth, scale=0.7]
 	
 	\draw [fill = black](8,0) circle (3pt);
 	\draw [fill = black](9,0) circle (3pt);
 	\draw [fill = black](8.5,0.87) circle (3pt);
 	\draw [fill = black](8.5,-0.87) circle (3pt);
        
    \draw[->-,ultra thick] (8.5,0.87) -- (9,0);
    \draw[->-, ultra thick] (8,0) -- (8.5,-0.87);
    \draw[densely dotted] (8,0) -- (8.5,0.87);
    \draw[densely dotted] (9,0) -- (8.5,-0.87);
    \draw[->-,thick] (8.5,-0.87) .. controls (7,-1) and (7,1) .. (8.5,0.87);
    \draw[->-,thick,densely dotted] (9,0) -- (8,0);
	\end{tikzpicture}
\end{equation*}
With open boundary conditions, such configurations can only arise when the transition loop connects the two dimers directly via the interior bond of the plaquette, which in turn implies that the interior bond belongs to the reference dimer configuration. (Notice that with periodic boundary conditions this assertion no longer holds. The transition loop can wind around a non-contractible cycle to accommodate two dimers with parallel Kasteleyn arrows).
In this case we have $i\gamma_{p,1}\gamma_{p,2}\ket{\mathbf{0}}= \pm\ket{\mathbf{0}}$ by definition and hence $\bra{\mathbf{0}}\mathcal{B}_{p} \ket{F(D)}=e^{\mp i\theta_p}e^{\pm i \pi/4}\braket{\mathbf{0}|F(D)}$.  The additional phase factor $e^{\pm i \pi/4}$ is exactly cancelled by our choice of $\theta_p=\pi/4$, and again we have $ \bra{\mathbf{0}}\mathcal{B}_{p} \ket{F(D)}=\braket{\mathbf{0}|F(D)}>0$.

\end{enumerate}

We have now demonstrated that Eq.~\eqref{BpResult} holds for systems with open boundary conditions.  Inserting this relation into Eq.~\eqref{hApp} immediately yields $h_{DD'}=-\braket{F(D')|\mathcal{B}_p|F(D)}=-1$ (again with open boundary conditions) whenever $D$ is flippable to $D'$ by $B_p$. So the corresponding dimer model is unfrustrated.  
 
We would like to remark that the proof here does not rely on the specific geometry of the lattice in an essential way, and can be readily adapted to Majorana-dimer flips on tetragonal plaquettes in other lattices provided one keeps track of the phase factor appearing in the last case.  Appendix~\ref{CommutationApp} describes a procedure for adapting these tools to the Fisher lattice.

\section{Fermionic Plaquette Operators on the Fisher Lattice}
\label{sec:proof-plaq}

\subsection{Matrix Elements of the Fermionic Plaquette Operator}
Here we will show that the matrix elements of $\mathcal{B}_p^{\varhexagon}$, again defined through
\begin{equation}
	\mathcal{B}_p^{\varhexagon}\ket{F(D)}=U_{1,2n-1}\cdots U_{1,5}U_{1,3}\ket{F(D)},
	\label{BpApp}
\end{equation}
indeed conform to Eq.~\eqref{eqn:overlap_Bp} as claimed in the main text.  
It suffices to focus only on Majoranas within a loop that is cycled by $\mathcal{B}_p^{\varhexagon}$.
Consider such a loop in the initial configuration $D$ that contains $2n$ Majoranas paired up as $is_{2j-1,2j}\gamma_{2j-1}\gamma_{2j}=1$ where $j=1, \dots, n$. Defining parity projectors 
\begin{equation}
  P_{i,j}=\frac{1+is_{ij}\gamma_i\gamma_j}{2},
\end{equation}
we then have 
\begin{equation}
	P_{2j-1,2j}\ket{F(D)}=\ket{F(D)}, \forall~j.
	\label{}
\end{equation}
A plaquette move initiated by $\mathcal{B}_p^{\varhexagon}$ (and its bosonic-sector counterpart $B_p^{\varhexagon}$) sends $D\rightarrow D'$ and yields a new Majorana dimerization pattern with
\begin{equation}
	P_{2j,2j+1}\ket{F(D')}= \ket{F(D')}, \forall~j.
	\label{FDprimeParity}
\end{equation}

Next we deduce the action of the braid operators $U_{i,j}$ in Eq.~\eqref{BpApp}.  Because $is_{12}\gamma_1\gamma_2=1$ when acting on $|F(D)\rangle$ (and using $s_{23} = s_{12} s_{13}$), we have
\begin{equation}
	\begin{split}
	U_{1,3}\ket{F(D)}&=\frac{1+s_{13}\gamma_1\gamma_3\cdot is_{12}\gamma_1\gamma_2}{\sqrt{2}}\ket{F(D)}\\
	&=\sqrt{2}P_{2,3}\ket{F(D)}.
	\end{split}
	\label{}
\end{equation}
After the exchange, the state $U_{1,3}\ket{F(D)}$ now has $is_{23}\gamma_2\gamma_3=1$ and $is_{14}\gamma_1\gamma_4=1$ owing to the projector $P_{2,3}$.
Iterating this procedure for the remaining braid operators yields
\begin{equation}
	\mathcal{B}_p^{\varhexagon}\ket{F(D)}=(\sqrt{2})^{n-1}P_{2n-2,2n-1}\cdots P_{4,5}P_{2,3}\ket{F(D)}.
	\label{}
\end{equation}
With the aid of Eq.~\eqref{FDprimeParity} we therefore immediately obtain
\begin{equation}
	\langle F(D')|\mathcal{B}_p^{\varhexagon}|{F(D)}\rangle = (\sqrt{2})^{n-1} \langle F(D')|F(D)\rangle.
	\label{eqn:bp_matrix_element_1}
\end{equation}
Finally, we note that while the phase of the overlap $\langle F(D')|F(D)\rangle$ is ambiguous, the norm is fixed:
\begin{equation}
	|\langle F(D')|F(D)\rangle|=2^{(1-n)/2}.
	\label{}
\end{equation}
This relation allows us to rewrite Eq.~\eqref{eqn:bp_matrix_element_1} in the desired form,
\begin{equation}
	\langle F(D')|\mathcal{B}_p^{\varhexagon}|{F(D)}\rangle=\frac{\langle F(D')|F(D)\rangle}{|\langle F(D')|F(D)\rangle|}.
	\label{eqn:bp_matrix_element_2}
\end{equation}

\subsection{Commutation Relations of Plaquette Operators}
\label{CommutationApp}
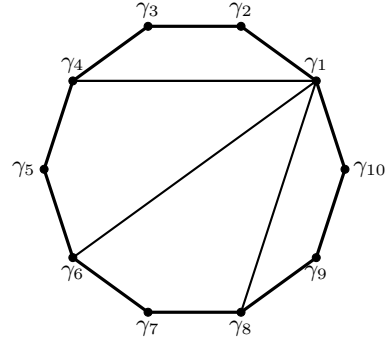
\begin{figure}
	\begin{tikzpicture}[>=stealth, scale=0.5]
	\foreach \i in { 1, 2, 3, 4, 5, 6, 7,8,9,10} {
		\draw[very thick] ({4*cos(0.2*180*\i)},{4*sin(0.2*180*\i)}) -- ({4*cos(0.2*180*\i+0.2*180)},{4*sin(0.2*180*\i+0.2*180)});
	}
	\foreach \i in { 1, 2, 3, 4} {
		\draw[fill=black] ({4*cos(0.2*180*\i)},{4*sin(0.2*180*\i)}) circle (3pt) node[anchor = south] {$\gamma_{\i}$};
	}
	\draw[fill=black] ({4*cos(0.2*180*5)},{4*sin(0.2*180*5)}) circle (3pt) node[left] {$\gamma_{5}$};
	\foreach \i in { 6, 7, 8, 9} {
		\draw[fill=black] ({4*cos(0.2*180*\i)},{4*sin(0.2*180*\i)}) circle (3pt) node[anchor=north] {$\gamma_{\i}$};
	}
	\draw[fill=black] ({4*cos(0.2*180*10)},{4*sin(0.2*180*10)}) circle (3pt) node[right] {$\gamma_{10}$};
	\draw[thick] ({4*cos(0.2*180*1)},{4*sin(0.2*180*1)}) -- ({4*cos(0.2*180*4)},{4*sin(0.2*180*4)});
	\draw[thick] ({4*cos(0.2*180*1)},{4*sin(0.2*180*1)}) -- ({4*cos(0.2*180*4)},{4*sin(0.2*180*4)});
	\draw[thick] ({4*cos(0.2*180*1)},{4*sin(0.2*180*1)}) -- ({4*cos(0.2*180*6)},{4*sin(0.2*180*6)});
	\draw[thick] ({4*cos(0.2*180*1)},{4*sin(0.2*180*1)}) -- ({4*cos(0.2*180*8)},{4*sin(0.2*180*8)});
	
	\end{tikzpicture}
	\caption{An example of tetragonalization for $t=4$.  A flip operator for the original 10-sided polygon may be decomposed into a series of elementary flips for each tetragon.  In this representation the tetragonal plaquettes, from top to bottom right, respectively correspond to $U_{1,3}$, $U_{1,5}$, $U_{1,7}$ and $U_{1,9}$ in Eq. \eqref{BpApp}.}
	\label{fig:tetragonalization}
\end{figure}

The goal of this section is to prove that $\mathcal{B}_p^{\varhexagon} \mathcal{B}_{p'}^{\varhexagon} = \mathcal{B}_{p'}^{\varhexagon} \mathcal{B}_{p}^{\varhexagon}$ in the restricted Hilbert space. 
As a primer it is very useful to first develop a geometric understanding of the operator $\mathcal{B}_p^{\varhexagon}$ by drawing an analogy to the plaquette operator in the triangular-lattice model. Imagine we partition the polygon enclosed by a transition loop into $t$ tetragons by connecting site $1$ with sites $4, 6, \dots, t-1$; see Fig.~\ref{fig:tetragonalization} for an illustration. One can view $\mathcal{B}_p^{\varhexagon}$ as implementing a series of elementary dimer flips through the tetragons as defined precisely as on the triangular lattice (first $[1,2,3,4]$, then $[1,4,5,6]$, and so on), provided we allow dimers to occupy the auxiliary edges at intermediate steps. The advantage of this `tetragonalization' is that we can easily track the phase of the fermionic state at each step (by looking at the overlap with some reference state) using the rules explained in Appendix \ref{sec:obc}.
	
With this geometric picture in hand, given a Majorana-dimer state $\ket{F(D)}$ one can write down a representation $\mathcal{B}_{p, T}^{\varhexagon}$ of the desired plaquette move acting on $\ket{F(D)}$ using some tetragonalization $T$ of the polygon.  In fact they are all equivalent in the restricted subspace, in the sense that for arbitrary tetragonalizations $T_{1}$ and $T_2$ and any dimer covering $D$, $\mathcal{B}_{p,T_{1}}^{\varhexagon}|F(D) \rangle =\mathcal{B}_{p,T_{2}}^{\varhexagon}|F(D) \rangle$.  One can see this as follows.  By construction, $\mathcal{B}_{p,T}^{\varhexagon}|F(D) \rangle$ gives a state corresponding to the same pairing of Majoranas for any tetragonalization $T$. Two tetragonalizations $T_1$ and $T_2$ thus generically give $\mathcal{B}_{p,T_{1}}^{\varhexagon}|F(D) \rangle = e^{i\phi_{p,D}} \mathcal{B}_{p,T_{2}}^{\varhexagon}|F(D) \rangle$.  One can conveniently isolate the phase factor on the right by taking an overlap with $|F(D)\rangle$: 
\begin{equation}
  e^{i \phi_{p,D}} = \frac{\langle F(D)| \mathcal{B}_{p,T_{1}}^{\varhexagon}|F(D) \rangle}{\langle F(D)| \mathcal{B}_{p,T_{2}}^{\varhexagon}|F(D) \rangle}.  
  \label{phipD}
\end{equation}
It turns out, however, that $\braket{F(D)|\mathcal{B}_{p,T}^{\varhexagon}|F(D)} >0$ independent of the tetragonalization $T$.  
We can view this expectation value as the overlap between $\mathcal{B}_{p,T}^{\varhexagon}|F(D)\rangle$ and a reference state $|F(D)\rangle$.  (Using this reference state instead of $|{\bf 0}\rangle$ is convenient here since the former more efficiently captures local effects of $\mathcal{B}_{p,T}^{\varhexagon}$.) 
In each elementary step, the tetragon dimer flip term with associated braid operator $U_{1, 2j+1}$ changes the number of loops in the corresponding transition graph---i.e., each step falls into either case 1 or 2 from Appendix~\ref{sec:obc}.  Thus the overlap with $\ket{F(D)}$ remains positive throughout so that $\braket{F(D)|\mathcal{B}_{p, T}^{\varhexagon}|F(D)}>0$ generically as claimed.  This property allows us to conclude that $\phi_{p,D}=0$ in Eq.~\eqref{phipD}, which in turn proves that $\mathcal{B}_{p,T_{1}}^{\varhexagon}|F(D) \rangle =\mathcal{B}_{p,T_{2}}^{\varhexagon}|F(D) \rangle$.  We can therefore safely drop the subscript $T$ hereafter.  The freedom of choosing any tetragonalization greatly simplifies the proof below.

We turn now to commutation of $\mathcal{B}_p^{\varhexagon}$'s in the restricted subspace.  Because the bosonic pieces of the flip term commute, one can readily see that $\mathcal{B}_p^{\varhexagon}\mathcal{B}_{p'}^{\varhexagon}|F(D)\rangle$ and $\mathcal{B}_{p'}^{\varhexagon}\mathcal{B}_{p}^{\varhexagon}|F(D)\rangle$ give states with identical Majorana pairing. In other words, these states at most differ by a complex phase factor.  One can show that the phases are also the same by analyzing the matrix elements $\braket{F(D)|\mathcal{B}_p^{\varhexagon}\mathcal{B}_{p'}^{\varhexagon}|F(D)}$ and $\braket{F(D)|\mathcal{B}_{p'}^{\varhexagon}\mathcal{B}_{p}^{\varhexagon}|F(D)}$, in a spirit similar to the proof in the previous paragraph.  While there are naively many different configurations to consider, several simplifications streamline the analysis.  First, we only need to consider the cases in which $p$ and $p'$ are neighboring plaquettes, since the commutation relation follows trivially otherwise.  Second, we can focus exclusively on the Majoranas that may be affected by both $\mathcal{B}_p^{\varhexagon}$ and $\mathcal{B}_{p'}^{\varhexagon}$, as shown in the following diagram:
\begin{equation*}
\begin{tikzpicture}[baseline={($ (current bounding box) - (0,3.0pt) $)},scale=1]
	\coordinate (a3) at ({cos(\PI/3)/3}, 0.5/3); 
	\filldraw  (a3) node[above] {$\gamma_{3}$} circle (0.05);
	\coordinate  (a2) at ({-cos(\PI/3)/3}, 0.5/3);
	\filldraw (a2) node[above] {$\gamma_{2}$} circle (0.05);
	
	\coordinate (a4) at ({4*cos(\PI/3)/3}, 4*0.5/3);
	\filldraw (a4) node[above]{$\gamma_{4}$} circle (0.05);
	
	\coordinate (a1) at ({-4*cos(\PI/3)/3}, 4*0.5/3);
	\filldraw (a1) node[above]{$\gamma_{1}$} circle (0.05);
	
	\coordinate (a9) at ({cos(\PI/3)/3}, -2-0.5/3);
	\filldraw (a9) node[below]{$\gamma_{9}$} circle (0.05);
	
	\coordinate (a8) at ({-cos(\PI/3)/3}, -2-0.5/3);
	\filldraw (a8) node[below]{$\gamma_{8}$} circle (0.05);
	
	\coordinate (a10) at ({4*cos(\PI/3)/3}, -2-4*0.5/3);
	\filldraw (a10) node[below]{$\gamma_{10}$} circle (0.05);
	
	\coordinate (a7) at ({-4*cos(\PI/3)/3}, -2-4*0.5/3);
	\filldraw (a7) node[below]{$\gamma_{7}$} circle (0.05);
	
	\coordinate (a5) at (0, -1/3);
	\filldraw (a5) node[right]{$\gamma_{5}$} circle (0.05);
	
	\coordinate (a6) at (0, -1-2/3);
	\filldraw (a6) node[right]{$\gamma_{6}$} circle (0.05);
	
	\draw[->-, thick] (a3) -- (a2);
	\draw[->-, thick] (a2) -- (a5);
	\draw[->-, thick] (a5) -- (a3);
	
	\draw[->-, thick] (a6) -- (a8);
	\draw[->-, thick] (a8) -- (a9);
	\draw[->-, thick] (a9) -- (a6);
	
	\draw[->-, thick] (a5) -- (a6);
	\draw[->-, thick] (a7) -- (a8);
	\draw[->-, thick] (a10) -- (a9);
	\draw[-<-, thick] (a1) -- (a2);
	\draw[-<-, thick] (a4) -- (a3);
	
	\draw ({-4*cos(\PI/3)/3}, -1) node {$p$};
	\draw ({4*cos(\PI/3)/3}, -1) node {$p'$};
\end{tikzpicture}\:.
\end{equation*}
And finally, it suffices to check only four different types of neighboring plaquette configurations.  For each one we tetragonalize the plaquette operators and keep track of the phase factors that arise.

\begin{figure}[t!]
\includegraphics[width=\columnwidth]{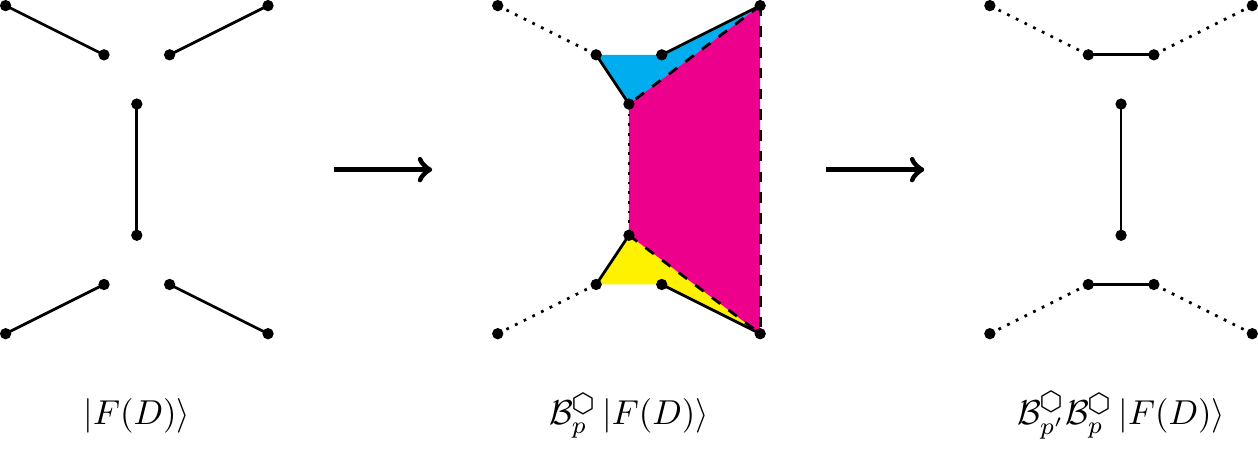}
\caption{Illustration of $\mathcal{B}_{p'}^{\varhexagon}\mathcal{B}_{p}^{\varhexagon}$ acting on an initial configuration without any loops in the overlap area between neighboring plaquettes $p$ and $p'$. The dashed lines in the middle figure are auxiliary lines for tetragonalization.}
\label{fig:commu1}
\end{figure}

To see how the proof works, consider an initial configuration $\ket{F(D)}$ where neither $p$ nor $p'$ has any loop extending in the overlapping region. Figure~\ref{fig:commu1} illustrates $\mathcal{B}_{p'}^{\varhexagon}\mathcal{B}_{p}^{\varhexagon}$ for this case. After applying $\mathcal{B}_p^{\varhexagon}$, the overlap with $\ket{F(D)}$ is positive as shown in the previous subsection. Then for $\mathcal{B}_{p'}^{\varhexagon}$ we tetragonalize $p'$ as indicated by the shaded regions in Fig.~\ref{fig:commu1}. As we can see from the illustration, the dimer flip at each step changes the transition loop number (again corresponding to case $1$ or $2$ from Appendix~\ref{sec:obc}), and therefore the overlap with $\ket{F(D)}$ remains positive. The reverse ordering $\mathcal{B}_{p}^{\varhexagon}\mathcal{B}_{p'}^{\varhexagon}$ works very similarly: After applying $\mathcal{B}_{p'}^{\varhexagon}$, one can tetragonalize $p$ such that each elementary dimer flip in $\mathcal{B}_{p}^{\varhexagon}$ changes the loop number. So we have shown that both $\braket{F(D)|\mathcal{B}_p^{\varhexagon}\mathcal{B}_{p'}^{\varhexagon}|F(D)}$ and $\braket{F(D)|\mathcal{B}_{p'}^{\varhexagon}\mathcal{B}_{p}^{\varhexagon}|F(D)}$ are positive.

\begin{figure}[t!]
\includegraphics[width=\columnwidth]{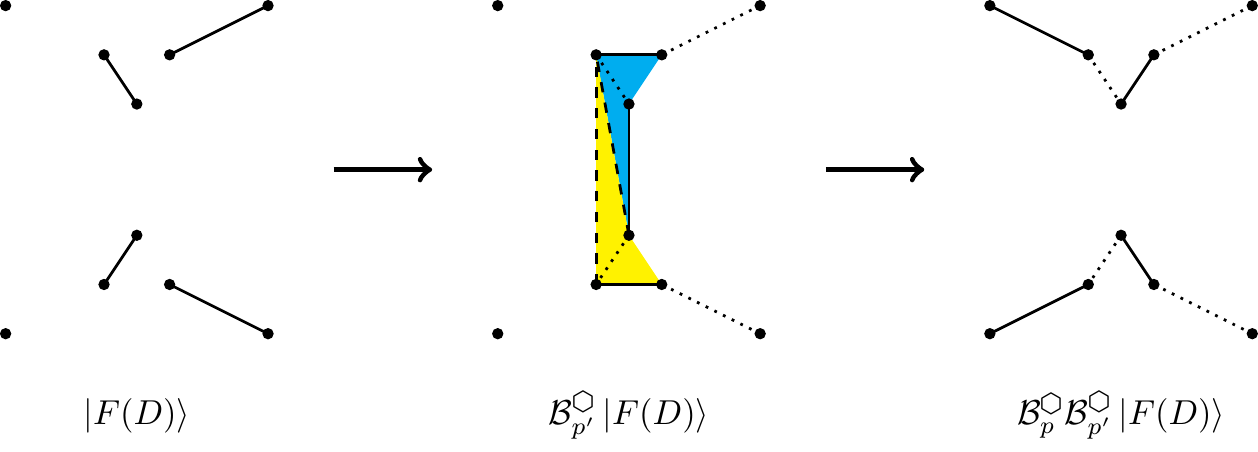}
\caption{Variation of Fig.~\ref{fig:commu1} in which the plaquette $p$ is occupied by a loop.  }
\label{fig:commu2}
\end{figure}

Next we consider a slightly more complicated initial configuration $\ket{F(D)}$ in which plaquette $p$ is occupied by a loop; for an illustration of 
$\mathcal{B}_{p}^{\varhexagon}\mathcal{B}_{p'}^{\varhexagon}$ here see Fig.~\ref{fig:commu2}. After applying $\mathcal{B}_{p'}^{\varhexagon}$ the overlap with $\ket{F(D)}$ is positive as usual. However, when we then apply $\mathcal{B}_{p}^{\varhexagon}$, some of the elementary dimer flips fall into case $3$ from Appendix~\ref{sec:obc} (see the two shaded tetragons in the middle figure).  It is then essential to carefully track the phases accumulated. It turns out that the phases cancel so that the overlap with $|F(D)\rangle$ remains positive in the end. For the opposite ordering $\mathcal{B}_{p'}^{\varhexagon}\mathcal{B}_{p}^{\varhexagon}$ one can tetragonalize without running into case 3, yielding $\braket{F(D)|\mathcal{B}_{p'}^{\varhexagon}\mathcal{B}_{p}^{\varhexagon}|F(D)}>0$ as well.  

The remaining two cases arise when the loop extends to both $p$ and $p'$ plaquettes beyond the overlapping region. By applying the same technique, one can see that for those cases both $\braket{F(D)|\mathcal{B}_p^{\varhexagon}\mathcal{B}_{p'}^{\varhexagon}|F(D)}$ and $\braket{F(D)|\mathcal{B}_{p'}^{\varhexagon}\mathcal{B}_{p}^{\varhexagon}|F(D)}$ encounter one elementary dimer flip that falls into case $3$ and that $\braket{F(D)|\mathcal{B}_{p'}^{\varhexagon}\mathcal{B}_{p}^{\varhexagon}|F(D)} = \braket{F(D)|\mathcal{B}_p^{\varhexagon}\mathcal{B}_{p'}^{\varhexagon}|F(D)} =|A|e^{\pm i \frac{\pi}{4}}$. 

Putting these results together, we see that $\mathcal{B}_p^{\varhexagon} \mathcal{B}_{p'}^{\varhexagon} = \mathcal{B}_{p'}^{\varhexagon} \mathcal{B}_{p}^{\varhexagon}$ in the restricted subspace, which is a key ingredient for obtaining a commuting-projector Hamiltonian on the Fisher lattice.  
 
\bibliography{ref}
\end{document}